\documentclass{aa} 

\usepackage{graphicx}
\usepackage{txfonts}
\usepackage{lipsum}
\usepackage{booktabs}
\usepackage{subcaption}         
\usepackage{lscape}             
\usepackage{float}              
\usepackage{placeins}           
\usepackage[colorlinks=true,citecolor=blue,linkcolor=blue,urlcolor=blue]{hyperref}
\usepackage[all]{hypcap}
\usepackage{xcolor}
\usepackage[utf8]{inputenc}
\usepackage{color}

\def\hi{\ifmmode {\mbox H{\scshape i}}\else H{\scshape i}\fi\xspace}
\def\hii{\ifmmode {\mbox H{\scshape ii}}\else H{\scshape ii}\fi\xspace}
\def\h2{\ifmmode {\mbox H$_2$}\else H$_2$\fi\xspace}

\begin{document}

   \title{ALMA Chemical Evolution (ACE) Survey: Molecular gas properties of low-mass, low-metallicity galaxies at cosmic noon}

   \titlerunning{ACE: Molecular gas properties of low-mass, low-metallicity galaxies at cosmic noon}
   \authorrunning{G. Popping et al.}

    \author{Gerg\H{o} Popping\inst{1}\fnmsep\thanks{Corresponding author: gpopping@eso.org}
        \and 
        Ivanna Langan\inst{2}
        \and
         Nikki N. Geesink\inst{1}
         \and
         Melanie Kaasinen\inst{3}
         \and
         Irene Shivaei\inst{2}
         \and
         Leindert A. Boogaard\inst{4}
         \and
         Daizhong Liu\inst{5,6}
         \and
         Max Parente\inst{7}
         \and
         Alexandra Pope\inst{8}
         \and
         Roxana Popescu\inst{8}
         \and
         Manuel Solimano\inst{2}
         \and
         Bahram Mobasher\inst{9} 
         \and 
         Bianca Moreschini\inst{1,10,11}
         \and
         Desika Narayanan\inst{7,12}
         \and
         Naveen Reddy\inst{13}
         \and
         Ryan L. Sanders\inst{14}
         }

   \institute{European Southern Observatory, Karl-Schwarzschild-Str. 2, D-85748, Garching, Germany
          \and
              Centro de Astrobiología (CAB), CSIC-INTA, Ctra. de Ajalvir km 4, Torrejón de Ardoz, E-28850, Madrid, Spain
        \and
              Research School of Astronomy and Astrophysics, Australian National University, Canberra, ACT 2611, Australia
        \and
              Leiden Observatory, Leiden University, PO Box 9513, NL-2300 RA Leiden, The Netherlands
        \and 
             Purple Mountain Observatory, Chinese Academy of Sciences, 10 Yuanhua Road, Nanjing 210023, China
       \and 
            State Key Laboratory of Radio Astronomy and Technology, Purple Mountain Observatory, Chinese Academy of Sciences, 10 Yuanhua Road, Nanjing 210023, China
        \and
            Department of Astronomy, University of Florida, 211 Bryant Space Sciences Center, Gainesville, FL 32611 USA
        \and
            Department of Astronomy, University of Massachusetts, Amherst, MA 01003, USA
        \and
             Department of Physics and Astronomy, University of California, Riverside, 900 University Avenue, Riverside, CA 92521, USA   
        \and
            Universit\'a di Firenze, Dipartimento di Fisica e Astronomia, via G.
Sansone 1, 50019 Sesto Fiorentino, Florence, Italy
        \and
            INAF – Arcetri Astrophysical Observatory, Largo E. Fermi 5, I-
50125, Florence, Italy
        \and
            Cosmic Dawn Center at the Niels Bohr Institute, University of Copenhagen and DTU-Space, Technical University of Denmark
        \and
            Department of Physics and Astronomy, University of California, Riverside, 900 University Avenue, Riverside, CA 92521, USA 
        \and
            Department of Physics and Astronomy, University of Kentucky, 505 Rose Street, Lexington, KY 40506, USA
             }


 
  \abstract
   {
Molecular gas plays a central role in regulating star formation and galaxy evolution, yet observational constraints at cosmic noon remain biased toward massive, metal-rich systems. We present ALMA Band~3 observations of the CO $J=3$--2 transition in 26 unlensed star-forming galaxies from the ALMA Chemical Evolution (ACE) Large Program at $z\sim2$--2.5, probing stellar masses of $10^{9} \lesssim M_\star \lesssim 10^{10.5}\,\mathrm{M}_{\odot}$ and sub-solar metallicities ($8.2 < 12 + \log(\mathrm{O/H}) < 8.6$). We derived molecular gas masses using a metallicity-dependent CO-to-H$_2$ conversion factor and alongside analysed stacking measurements and a homogenized literature compilation spanning both local and high-redshift galaxies. We find that the ACE galaxies extend established molecular-gas scaling relations to an order of magnitude lower stellar masses than previously explored at cosmic noon. The molecular gas mass ($M_{\rm mol}$) correlates tightly with star formation rate (SFR), while molecular gas fractions show a strong dependence on specific star formation rate (sSFR) and offset from the star-forming main sequence. In contrast, molecular gas fractions show only weak trends with stellar mass and no significant dependence on metallicity. Molecular gas depletion times are $\sim1\,\mathrm{Gyr}$ and vary little with stellar mass or metallicity, and weakly with sSFR and offset from the star-forming main sequence. Together this further reinforces that the availability of molecular gas is the primary driver of the SFR in galaxies, with changes in star-formation efficiency playing a secondary role. Leveraging the expanded parameter space probed by the ACE and literature samples, we derive a new empirical prescription for predicting $M_{\rm mol}$ as a function of SFR and sSFR. The persistence of the observed scaling relations across a broader range of galaxy properties than probed before at cosmic noon suggests a largely universal framework governing the molecular gas--star formation cycle across cosmic time.
  }


   \maketitle
   \nolinenumbers 
%

\section{Introduction}
\label{sec:intro}

Molecular gas is the fuel that drives star formation and galaxy evolution across cosmic time. Understanding the mechanisms that regulate how molecular gas is converted into stars remains a key goal in astrophysics. In the framework of the baryon cycle through galaxies, a galaxy's star formation rate ($\mathrm{SFR}$) is determined by a continuous balance between gas accretion from the cosmic web, star formation, and stellar-driven outflows \citep[e.g.,][]{Bouche2010, Dave2012, Lilly2013}. Within this paradigm, the molecular gas reservoir acts as the central engine. Scaling relations that link global galaxy properties -- like stellar mass ($M_{\star}$, star-formation rate (SFR), and redshift (z) -- to molecular gas properties -- like molecular gas mass ($M_{\mathrm{mol}}$), molecular-to-stellar mass ratio ($M_{\mathrm{mol}}/M_{\star}$), gas depletion time ($t_{\mathrm{dep}} \equiv M_{\mathrm{mol}}/\mathrm{SFR}$) -- offer a direct look into the physics regulating this cycle, showing how galaxies grow along the star-forming main sequence \citep[SFMS;][]{Noeske2007, Elbaz2010}. 

The redshift range $z \sim 1$--$3$ represents a particularly significant epoch in the history of the Universe, corresponding to the peak of cosmic star formation activity, often referred to as ``cosmic noon'' \citep{Madau2014}. Galaxies during this period formed stars at rates substantially higher than those observed in the local Universe. Massive galaxies had elevated gas fractions, enhanced star-formation efficiencies, and distinct interstellar medium (ISM) conditions, highlighting the need for direct observations of representative cosmic noon galaxy populations \citep{Steidel2014,Kaasinen2017,Tacconi2018,Aravena2019}. Testing the scaling relations that connect cold-gas content to stellar mass assembly at these redshifts is therefore essential for understanding the mechanisms that regulate galaxy growth and drive the peak of cosmic star formation.

Over the past decade, observational surveys with the Atacama Large Millimeter/submillimeter Array (ALMA), Plateau de Bure Interferometer (PDBI), and Northern Extended Millimeter Array (NOEMA) have revolutionized our view of the molecular gas in cosmic noon galaxies. By using both  $\mathrm{CO}$ line transitions and dust continuum emission as gas tracers, these surveys have established robust empirical scaling relations for the molecular gas content of galaxies across a range of redshifts, stellar masses, and SFRs \citep[e.g.,][]{Tacconi2013, Genzel2015, Scoville2017, Tacconi2018, Aravena2019, Liu2019, Tacconi2020, Gomez2022, Sanders2023}. These relations provide relevant constraints for the baryon cycle and indicate that a galaxy's location relative to the SFMS is tightly coupled to the availability of cold molecular gas. 

The star formation efficiency ($\mathrm{SFE}$) of galaxies is fundamentally regulated by localized physical conditions, rooted in cloud-scale densities and turbulent pressure and feedback from young stars and supernovae \citep{Ostriker2022}. These cloud properties, however, are shaped by the broader, large-scale dynamics of the galactic disk. This connection between small-scale star-forming environments and disk-scale structure is often parameterized in terms of gravitational stability via the Toomre $Q$ parameter \citep[e.g.,][]{Genzel2011,Wisnioski2015}. Ultimately, these localized variations scale up to determine the integrated gas depletion timescales ($t_{\mathrm{dep}} = M_{\rm mol} / \rm{SFR}$), which previous surveys have shown to evolve with cosmic time and with distance from the SFMS \citep{Aravena2019, Tacconi2020}. 

Despite this progress, current observational samples of unlensed field galaxies suffer from strong selection biases. Although dust continuum measurements are highly efficient for targeting large samples, $\mathrm{CO}$ line emission remains the most direct and physically robust way to trace the bulk molecular gas. However, existing unlensed $\mathrm{CO}$ observations at high redshift are heavily skewed toward massive ($M_{\star} \gtrsim 10^{10.5}\,\mathrm{M}_{\odot}$) and metal-rich (approximately solar metallicity) systems \citep[for recent reviews see][]{Hodge2020,Tacconi2020}. Because of this, we still lack a clear empirical picture of the molecular gas cycle in typical low-mass, low-metallicity galaxies at cosmic noon, even though these systems are the direct progenitors of Milky Way-like galaxies today. Moreover, most CO-observed galaxies lack the gas-phase metallicity measurements needed to robustly estimate  the CO-to-\h2 conversion factor \citep{Narayanan2012,Bolatto2013}.

To push toward low stellar masses ($M_{\star} \lesssim 10^{10.5}\,\mathrm{M}_{\odot}$) and low-metallicity ($12 + \log{(O/H)} < 8.6$) systems, observations have traditionally relied on the flux-boosting power of strong gravitational lensing \citep[e.g.,][]{Saintonge2013, DessaugesZavadsky2015, Motta2018, Solimano2022, Catan2024, Tsujita2025}. Although these pioneering lensing campaigns provide an invaluable look into the low-mass regime, they are intrinsically limited by small, stochastic samples that carry complex lens modelling and structural reconstruction uncertainties. Crucially, what is lacking is a statistically representative sample of \textit{unlensed} low-mass sources selected with clear, uniform criteria and backed by wide ancillary data, for which molecular gas \emph{and} metallicity information is available.  

The current lack of observations stands in contrast to theoretical models, which are already well-developed. Cosmological hydrodynamical simulations and semi-analytic models have advanced significantly, incorporating detailed treatments for the multi-phase ISM \citep[e.g.,][]{Obreschkow2009, Lagos2011, Popping2014, Popping2015, Lagos2015, Xie2017, Popping2019, Dave2020, Lagos2026}. These models make explicit predictions for the molecular gas properties of galaxies over cosmic time. They find that galaxies were more gas rich at cosmic noon than in the local universe and their molecular-to-stellar mass ratio decreases with increasing stellar mass \citep[e.g.,][]{Popping2015}. \citet{Lagos2026} furthermore finds a positive correlation between molecular gas depletion time and metallicity on resolved scales and that at fixed metallicities the depletion time increases with time. Testing these predictions requires expanding our observational boundaries into the low-mass, metal-poor regime using well-understood selection functions across cosmic time.

In this paper, we present the molecular gas content of galaxies observed as part of the ALMA Chemical Evolution (ACE) survey, an ALMA Large Program designed to bridge this observational gap. ACE targets the CO J$=$3--2 emission line in unlensed $z=2-2.5$ galaxies with stellar masses nearly an order of magnitude lower ($10^{9} < M_{\star} < 10^{10.5}\,\mathrm{M}_{\odot}$) and gas-phase metallicities roughly $0.4\,\mathrm{dex}$ lower ($8.2 < 12 + \log(\mathrm{O/H}) < 8.6$) than previous unlensed $\mathrm{CO}$ samples. Crucially, this program pairs these new $\mathrm{CO}$ observations with robust gas-phase metallicity measurements based on strong nebular emission lines and a wide suite of multi-wavelength ancillary data \citep{Shivaei2026}. This unique combination allows us to establish a clearer view of molecular gas scaling relations and how cold gas reservoirs shape galaxy assembly in a lower mass and lower metallicity regime than has been possible until now.

This paper is organized as follows. In Section~\ref{sec:data}, we present the ACE sample, the ALMA Band 3 CO J$=$3--2 observations, and the literature data used for comparison. In Section~\ref{sec:co-to-H2}, we describe our methods for measuring $\mathrm{CO}$ line fluxes and converting them into molecular gas masses. We present our results in Section~\ref{sec:analysis}, discuss their implications in Section~\ref{sec:discussion}, and summarize our main findings in Section~\ref{sec:summary}. Throughout this work, we assume a \citet{Chabrier2003} initial mass function (IMF) and adopt a \citet{Planck2020} cosmology with $H_0 = 67.66\,\mathrm{km}\,\mathrm{s}^{-1}\,\mathrm{Mpc}^{-1}$, $\Omega_{\mathrm{m}} = 0.3111$, and $\Omega_{\Lambda} = 0.6889$. Metallicity is defined in terms of the gas-phase oxygen abundance, $12 + \log(\mathrm{O}/\mathrm{H})$, and throughout the paper we adopt a solar metallicity of $12 + \log(\mathrm{O}/\mathrm{H})  _\odot = 8.69$ \citep{Asplund2009}. All presented molecular gas masses include a contribution from helium and metals.

\section{Sample and Observations}
\label{sec:data}
\subsection{ACE sample selection and properties}
\label{subsec:sample}
The primary sample analyzed in this work is drawn from the ALMA large program ACE (2024.1.00534.L), which targets a subset of the star-forming galaxies (SFGs) at $z \sim 2.0\text{--}2.5$, previously characterized in \citet{Shivaei2022}. These sources were originally selected from the MOSFIRE Deep Evolution Field (MOSDEF) survey \citep{Kriek2015}, comprising $\sim 1500$ galaxies with Keck/MOSFIRE near-infrared (NIR) spectra. \citet{Shivaei2022} presented $1.2\,\mathrm{mm}$ dust continuum measurements for $27$ typical main-sequence SFGs. The ACE large program has built a sample of $25$ of these galaxies with measurements of the CO J$=$3--2 emission line, $3\,\mathrm{mm}$, and $1\,\mathrm{mm}$ dust continuum emission. We use the 22 ACE galaxies targeted in ALMA band 3 (the remaining 3 galaxies being only targeted in ALMA band 7) and an additional 4 galaxies with ALMA band 3 CO J$=$3--2 detections from \citet{Sanders2023}, totaling 26 galaxies (one on top of the ACE sample of 25, but with the exact same available ancillary information).
Because the ACE sample is located within the Cosmic Evolution Survey (COSMOS) field \citep{Scoville2007}, we leverage the extensive multi-wavelength photometry available (e.g., from the 3D-HST survey; \citealt{Skelton2014}), supplemented by new \textit{James Webb} Space Telescope (\textit{JWST}) NIRCam and MIRI data from the COSMOS2025 \citep{Casey2023, Shuntov2025} and PRIMER \citep{Dunlop2021} programs. Using this photometric baseline and our ALMA continuum observations \citep{Popescu2026}, in \citet{Shivaei2026} we derive stellar masses ($M_\star$) and $\mathrm{SFR}$s averaged over $10\,\mathrm{Myr}$ via spectral energy distribution (SED) fitting with the \textsc{Prospector} code \citep{Leja2017, Johnson2021}. We assume a delayed-$\tau$ star formation history with a flat age prior between 1 Myr and the age of the universe at the redshift of the galaxy. In \citet{Shivaei2026} we furthermore determined gas-phase metallicities from multiple rest-frame optical and NIR emission lines (spanning $[\mathrm{OII}]\lambda\lambda 3726, 3729$ to $[\mathrm{SII}]\lambda\lambda 6716, 6731$; \citealt{Kriek2015}) adopting the strong-line calibrations of \citet{Sanders2025}. The final sample spans a range of $9.13 < \log(M_*/\mathrm{M}_\odot) < 10.62$ and $1.3 < \log(\mathrm{SFR}/\mathrm{M}_\odot\,\mathrm{yr}^{-1}) < 2.2$ and metallicities from $12 + \log(\mathrm{O/H}) = 8.19$ to 8.61. The derived properties of the ACE targets are listed in Table~\ref{tab:ace_properties}. More details on the SED-fitting, metallicity estimation, and sample characteristics are provided in \citet{Shivaei2026}. 

We note that other ACE studies have adopted star-formation rates (SFRs) derived from H$\alpha$ emission and corrected for dust attenuation using the Balmer decrement. In this work, however, we use SFRs obtained from spectral energy distribution (SED) fitting (see \citealt{Shivaei2026} for the SED fitting details), as H$\alpha$-based SFR measurements are not available for the literature comparison sample (Section~\ref{subsec:literature}). Adopting SED-based SFRs therefore enables the most consistent comparison between the two samples.

\begin{figure*}
  \centering
  \includegraphics[width=\textwidth]{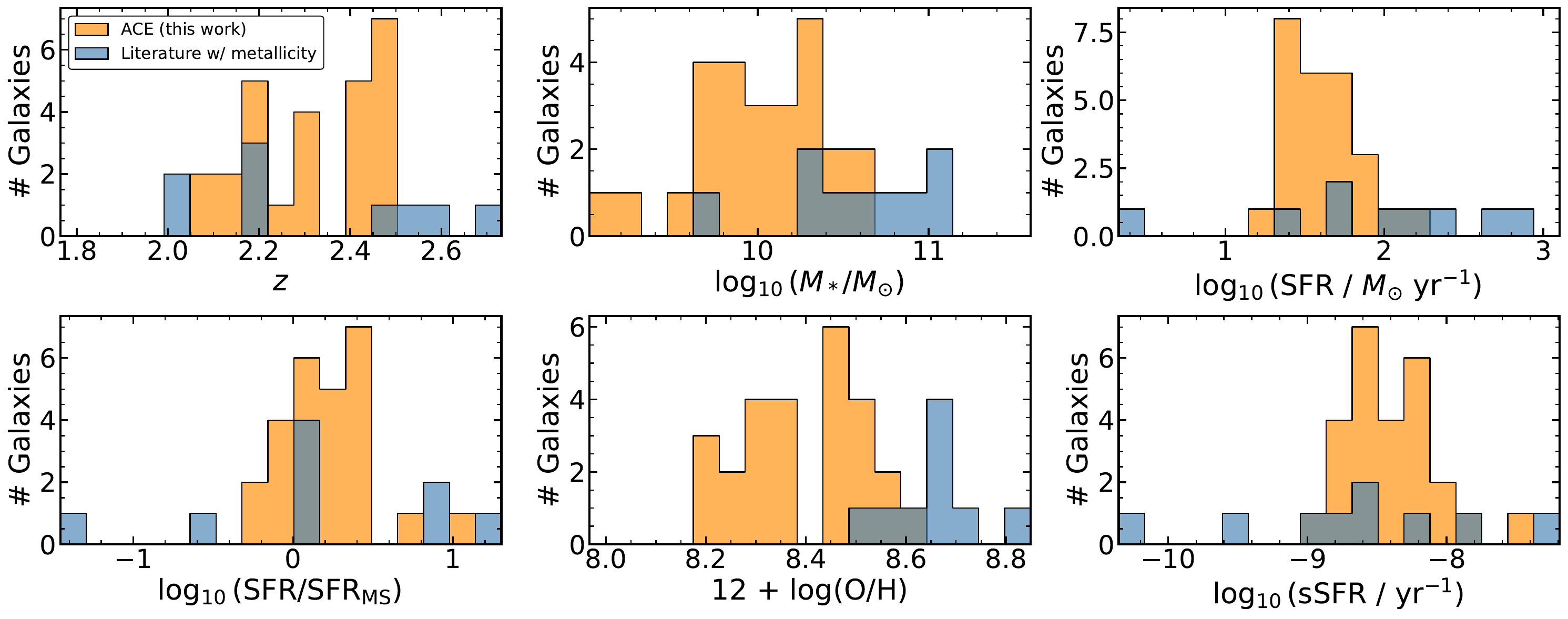}
  \caption{The distribution of galaxy properties (redshift, stellar mass, $\mathrm{SFR}$, $\mathrm{SFR}$ relative to the main sequence from \citet{Popesso2023}, metallicity, and $\mathrm{sSFR}$) for the ACE targets and the $z\sim2-2.5$ literature compilation for which CO and metallicity information is available. The ACE targets probe a mostly unexplored regime in stellar mass, metallicity, and $\mathrm{SFR}$.
  \label{fig:property_histograms}}
\end{figure*}

\subsection{ALMA Band 3  CO J$=$3--2  observations}
The ALMA Band 3 CO J$=$3--2 observational setup, data calibration, imaging, and flux measurement strategies are presented in detail in \citet{Langan2026} and \citet{Shivaei2026}. Below, we briefly summarize the adopted approaches, but we refer the reader to the aforementioned papers for more detailed descriptions. The described approaches were applied to the targets observed as part of the ACE Large Program, as well as the archival datasets originally presented in \citet{Sanders2023}. 

We make use of calibrated measurement sets provided by the European ALMA Regional Centre through the calMS service \citep{Petry2020}. For the imaging of the data, we make use of the {Common Astronomy Software Applications} ({CASA} v6.5.21; \citealt{CASA}) package. We use the {\tt tclean} task with natural weighting ($\mathtt{robust=2}$) to create cubes, with a channel width of $\Delta v = 7.8125\,\mathrm{MHz}$ ($\sim 23\,\mathrm{km\,s}^{-1}$). This provides an optimal balance between the signal-to-noise ratio ($S/N$) and the sampling of the line profile. The aim of the ACE program is to fully recover the CO J$=$3--2 line flux of the sources. To this end, we generated data cubes at the native resolution as well as with $1\arcsec$ and $2\arcsec$ $uv$-tapering. We next defined a scoring function consisting of the product of the integrated flux and the peak $S/N$ for each image to evaluate which of these images is best suited to recover the total flux of the source within a single beam element. The image with the highest score (among the native resolution, $1\arcsec$, and $2\arcsec$ tapering, respectively) was used for subsequent analysis. Each selection was visually inspected against \textit{JWST} imaging to prevent flux contamination from neighboring sources. We refer the reader to \citet{Langan2026} for a more detailed description. 

The derivation of optimized CO J$=$3--2 moment-0 images and spectra followed an iterative procedure also presented in \citet{Langan2026}. Initially, we collapsed the data cubes over a velocity range corresponding to the H$\alpha$ full-width-half-maximum \citep{Kriek2015} around the systemic redshift. From this preliminary image, we identified the emission peak and extracted a 1D spectrum, which was then fitted to determine the $\mathrm{CO}$ line width. This width was subsequently used to define the velocity interval for the final, optimized moment-0 image. As a final step, we extracted the CO J$=$3--2 spectrum from the peak pixel of this optimized image.

\subsection{Literature datasets}
\label{subsec:literature}
\subsubsection{$z=2-2.5$ literature compilation with metallicity and CO information}
We complement our ACE sample with literature data at similar redshifts from the Plateau de Bure High-$z$ Blue Sequence Survey \citep[PHIBSS;][]{Tacconi2013, Tacconi2018} and ALMA Spectroscopic Survey in the Hubble Ultra Deep Field \citep[ASPECS;][]{Aravena2019, GonzalezLopez2019, Boogaard2019, Boogaard2020}. These literature galaxies are selected based on the availability of molecular gas measurements derived from $\mathrm{CO}$ observations, as well as the availability of optical strong lines observed through either ground-based or JWST/NIRSpec spectroscopy (after cross-matching to the DAWN JWST Archive spectroscopic release; \citealt{Valentino2025,Pollock2026}), alongside complementary JWST/NIRCam coverage. This results in 8 galaxies from ASPECS and 6 galaxies from PHIBSS. Nine of these galaxies roughly fall within our target redshift range of $z \sim 2$--$2.5$. To ensure a self-consistent analysis, we derived the $M_{\star}$ and $\mathrm{SFR}$ for these targets by applying the previously mentioned SED-fitting methodology explained in \citet{Shivaei2026}, and their metallicities using the \citet{Sanders2025} calibration adopted for the ACE galaxies. Molecular gas masses ($M_{\mathrm{mol}}$) are derived using the exact same assumption for the CO-to-\h2 conversion factor and CO(3--2) to CO(1--0) excitation correction as those applied to the ACE targets (see Section~\ref{sec:co-to-H2}). 

In many of the figures throughout this paper, we also include the compilation of galaxies at $z \sim 2$--$2.5$ with $\mathrm{CO}$-based molecular gas masses presented in the review by \citet[including lensed and unlensed galaxies]{Tacconi2020}. We adopt the molecular hydrogen masses exactly as presented in the \citet{Tacconi2020} compilation, which assume a constant $\mathrm{CO}$-to-$\mathrm{H}_2$ conversion factor of $\alpha_{\mathrm{CO}} = 4.6\,\mathrm{M}_{\odot}\,(\mathrm{K}\,\mathrm{km}\,\mathrm{s}^{-1}\,\mathrm{pc}^2)^{-1}$, as we lack metallicity information to rescale them to the \citet{Accurso2017} conversion factor. We include this compilation to place our results in the context of the extensive literature that has set the standard in the field; however, we note that the molecular gas masses derived in this work for the ACE sample and our new literature compilation are more robust, given that they account for the explicit dependence of the $\mathrm{CO}$-to-$\mathrm{H}_2$ conversion factor on measured metallicities.

\subsubsection{$z=0$ compilation} 
To place our findings in the context of local galaxies, we compare the ACE sample to the $z=0$ galaxy compilation from \citet{Galliano2021}, focusing specifically on sources with secure $\mathrm{CO}$ detections. This local sample spans a significantly wider metallicity range than alternative low-redshift compilations, making it an ideal baseline for comparison with the low-metallicity regime probed by ACE. Because a subset of sources in the \citet{Galliano2021} parent catalog lacked $\mathrm{CO}$-based molecular gas information, we cross-matched their compilation with the sample from \citet{Colombo2025} to supplement the missing $\mathrm{CO}$ fluxes where available. This cross-matching yields a final local comparison sample of 143 galaxies. For a consistent comparison, we recalculate the molecular gas masses for this local sample adopting the same $\mathrm{CO}$-to-$\mathrm{H}_2$ conversion factor methodology described in Section~\ref{sec:co-to-H2}.

\section{CO line flux measurement and molecular gas mass estimation}
\label{sec:co-to-H2}
The derivation of CO J$=$3--2 line properties and integrated fluxes is based on the Markov chain Monte Carlo (MCMC) Gaussian fitting procedure described in detail in \citet{Langan2026}. Briefly, we generated $5000$ noise realizations for each spectrum by perturbing the data with Gaussian noise scaled to the local $\mathit{rms}$ level. Each realization was fitted with a single Gaussian profile to determine the median integrated flux, $F_{\mathrm{int}}$, and the associated kinematic parameters. For sources with $\mathrm{S/N} < 4$, we repeated the fitting procedure by fixing the Gaussian width to the corresponding $\mathrm{H}\alpha$ line width ($\mathrm{FWHM}_{\mathrm{H}\alpha}$; \citealt{Kriek2015}), accounting for the respective spectral resolutions. 

We adopted a conservative estimate for the flux uncertainties, defined as the maximum of the standard deviation from the MCMC realizations and the noise-based estimate, $\mathit{rms}_{\mathrm{av}} \times \sqrt{N_{\mathrm{chan}}} \times \Delta v$, where $N_{\mathrm{chan}}$ is the number of channels within the line $1\sigma$ width. Non-detections are defined as sources with an integrated flux $\mathrm{S/N} < 3$. For these objects, we adopt $3\sigma$ upper limits calculated as $3 \times \mathit{rms}_{\mathrm{av}} \times \sqrt{N_{\mathrm{chan}}} \times \Delta v$, where $N_{\mathrm{chan}}$ corresponds to the velocity interval of $2 \times \mathrm{FWHM}_{\mathrm{H}\alpha}$. In total, our sample consists of $17$ detections with a mean $\mathrm{S/N} \approx 5$ and $9$ upper limits \citep{Shivaei2026}.  The extracted CO J$=$3--2 flux densities are listed in Table~\ref{tab:ace_properties}.

To derive molecular gas masses from the CO($3$--$2$) observations, we first convert the integrated line fluxes to CO($1$--$0$) luminosities ($L'_{\mathrm{CO}(1-0)}$) by adopting a line brightness temperature ratio of $r_{31} = L'_{\mathrm{CO}(3-2)}/L'_{\mathrm{CO}(1-0)} = 0.77 \pm 0.14$. This value is taken from \citet{Boogaard2020}, who derived it from a stacking analysis of typical star-forming galaxies at $z\sim2.5$. The adoption of this excitation correction is further supported by observations of local dwarf galaxies, which show no evidence for systematically different CO excitation conditions compared to more metal-rich systems \citep{Meier2001, Cormier2014}. Nevertheless, the CO excitation properties of low-mass, low-metallicity galaxies at high redshift remain largely unexplored observationally, and the applicability of this value in the regime probed by ACE has yet to be tested directly (see also \citealt{Anirudh2025} for theoretical considerations).

The choice of the $\mathrm{CO}$-to-$\mathrm{H}_2$ conversion factor ($\alpha_{\mathrm{CO}} = M_{\mathrm{mol}} / L'_{\mathrm{CO}(1-0)}$) is critical, as it is known to vary with gas-phase metallicity \citep[e.g.,][]{Narayanan2012,Bolatto2013}. To determine the most suitable CO-to-H$_2$ conversion factor for our sample, \citet{Langan2026} used the dynamical mass method to obtain independent estimates of $\alpha_{\mathrm{CO}}$ from the observed CO line widths. These values were subsequently compared with several metallicity-dependent $\alpha_{\mathrm{CO}}$ prescriptions from the literature \citep[e.g.,][]{Bolatto2013,Accurso2017,Madden2020}. \citet{Langan2026} found that the dynamical estimates are in  best agreement with the calibration presented in  \citet{Accurso2017}, which is subsequently the conversion used throughout the ACE papers.
The resulting molecular gas masses, which include a contribution for Helium, are presented in \citet{Langan2026} and for clarity also listed in Table~\ref{tab:ace_properties}.


\subsection{Stacking}
Despite the depth of our data, for nine sources the CO J$=$3--2 emission remains undetected. We use a cube stacking analysis to recover average sample properties benefitting from the combined depth of the ALMA band 3 observations. We follow the methodology outlined in \citet{Geesink2026} to obtain stacked fluxes in discrete bins of stellar mass ($M_{\star}$), metallicity, SFR, specific SFR (sSFR), and the distance from the main sequence ($\text{SFR}/\text{SFR}_{\text{MS}}$). The resulting stacked CO fluxes and the corresponding molecular gas masses derived from these detections are summarized in Table~\ref{tab:stacks}. We point the reader to \citet{Geesink2026} for a more detailed description of the cube stacking approach, especially the error-handling.

\begin{figure}
  \centering
  \includegraphics[width=\columnwidth]{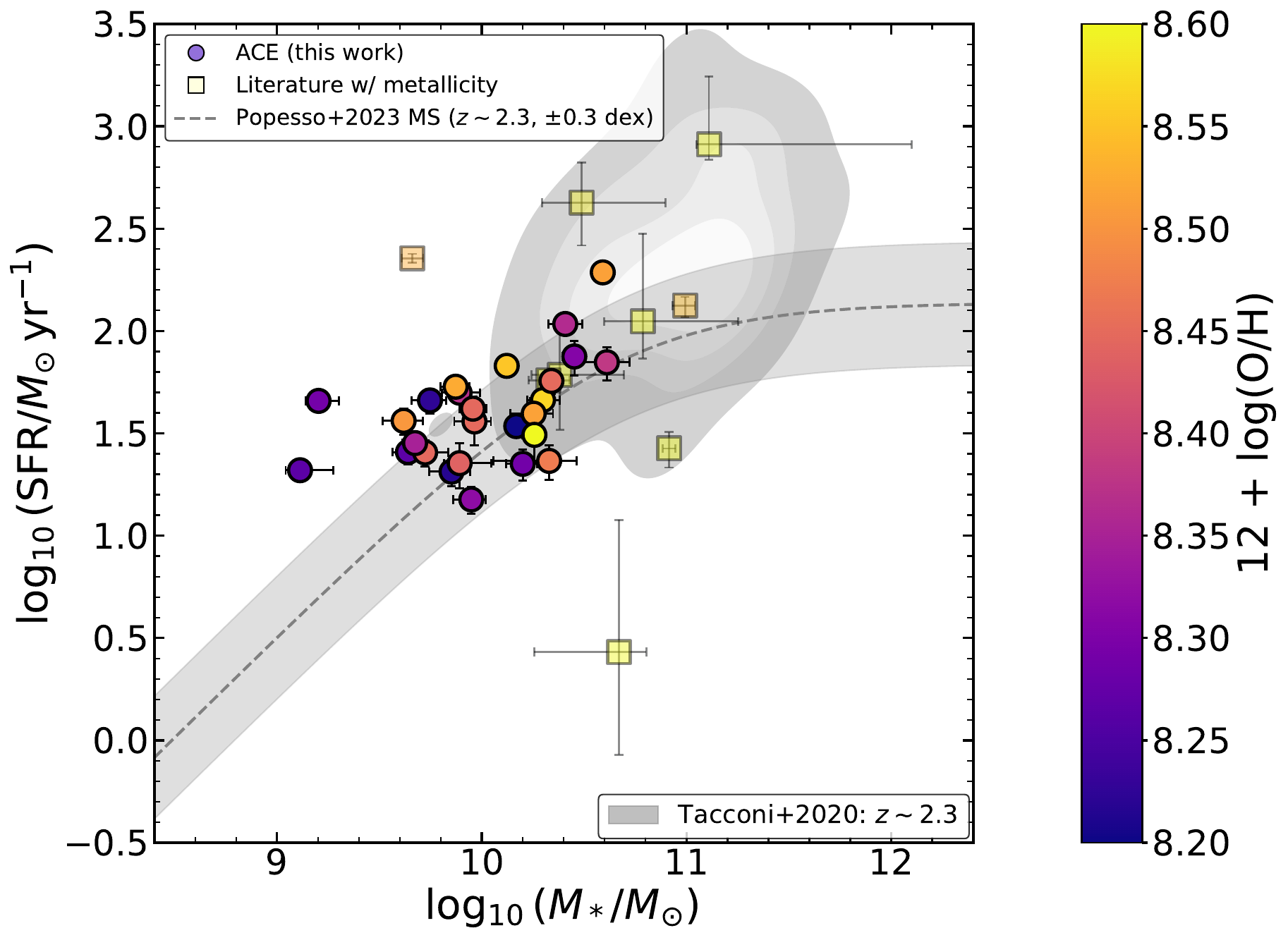}
  \caption{Position of the ACE galaxies relative to the SFMS, colour-coded by their gas-phase metallicity. Circles mark the ACE sample, while squares indicate galaxies from the literature compilation with available metallicity information. The grey shaded contours show the broader cosmic noon population from \citet{Tacconi2020}. For comparison, the dashed line tracks the empirical main sequence derived by \citet{Popesso2023}, with the surrounding shaded region representing a typical $0.3\,\mathrm{dex}$ scatter.}
  \label{fig:mstar_vs_sfr}
\end{figure}

\begin{table*}[h]
\centering
\caption{Censored Kendall rank correlation \citep{Akritas1996} for the ACE and $z\sim2$ literature compilation for the various relations presented in Figures~\ref{fig:Mstar_mmol_sfr}-\ref{fig:four_panel_tdep}. $\tau$ is Kendall's statistic (in the range -1 to 1) and $p$ is the two-sided $p$-value, with $p$-value indicating the probability of obtaining the observed correlation purely by random chance. The last column indicates the correlation strength, where we list all trends with an absolute Kendall correlation coefficient of $|\tau| > 0.30$ as Strong, $0.15 < |\tau|< 0.30$ as Weak, and $|\tau| < 0.15$ as None.}
\label{tab:kendall_censored_scaling}
\begin{tabular}{ccccc}
\hline\hline
$x$ & $y$ & $\tau$ & $p$ &  Correlation strength\\
\hline
$\log_{10}(M_{\star}/M_{\odot})$ & $\log_{10}(M_{\mathrm{mol}}/M_{\odot})$ & $0.259$ & $2.874\times 10^{-2}$&Weak\\
$\log_{10}(M_{\mathrm{mol}}/M_{\odot})$ & $\log_{10}(\mathrm{SFR}/M_{\odot}\,\mathrm{yr}^{-1})$ & $0.484$ & $4.313\times 10^{-5}$ & Strong\\
$\log_{10}(M_{\star}/M_{\odot})$ & $\log_{10}(M_{\mathrm{mol}}/M_{\star})$ & $-0.276$ & $1.986\times 10^{-2}$ & Weak\\
$12+\log_{10}(\mathrm{O/H})$ & $\log_{10}(M_{\mathrm{mol}}/M_{\star})$ & $-0.087$ & $4.602\times 10^{-1}$  & None\\
$\log_{10}(\mathrm{sSFR}/\mathrm{yr}^{-1})$ & $\log_{10}(M_{\mathrm{mol}}/M_{\star})$ & $0.360$ & $2.373\times 10^{-3}$ & Strong\\
$\log_{10}(\mathrm{SFR}/\mathrm{SFR}_{\mathrm{MS}})$ & $\log_{10}(M_{\mathrm{mol}}/M_{\star})$ & $0.336$ & $4.507\times 10^{-3}$ & Strong\\
$\log_{10}(M_{\star}/M_{\odot})$ & $\log_{10}(t_{\mathrm{dep}}/\mathrm{yr})$ & $0.000$ & 1 &None \\
$12+\log_{10}(\mathrm{O/H})$ & $\log_{10}(t_{\mathrm{dep}}/\mathrm{yr})$ & $-0.030$ & $7.982\times 10^{-1}$ &None\\
$\log_{10}(\mathrm{sSFR}/\mathrm{yr}^{-1})$ & $\log_{10}(t_{\mathrm{dep}}/\mathrm{yr})$ & $-0.155$ & $1.914\times 10^{-1}$ &Weak\\
$\log_{10}(\mathrm{SFR}/\mathrm{SFR}_{\mathrm{MS}})$ & $\log_{10}(t_{\mathrm{dep}}/\mathrm{yr})$ & $-0.155$ & $1.914\times 10^{-1}$ &Weak\\
\hline\hline
\end{tabular}
\end{table*}

\section{Analysis}
\label{sec:analysis} 
\subsection{ACE galaxies in context of literature compilations}
To place the ACE galaxies in the context of the available information in the literature, we first show the distribution of properties of the ACE targets compared to the $z \sim 2$--$2.5$ literature compilation in Figure~\ref{fig:property_histograms}. ACE has a relatively even distribution of metallicities between $12 + \log(\mathrm{O/H}) \sim 8.2$ and $8.6$. The ACE sample newly probes lower metallicities, $12 + \log(\mathrm{O/H}) < 8.5$, than has been achieved for unlensed targets at cosmic noon thus far. The ACE sample is mostly centered at $M_{\star} < 10^{10.5}\,\mathrm{M}_{\odot}$, with $\mathrm{SFR}$s between $10$ and $100\,\mathrm{M}_{\odot}\,\mathrm{yr}^{-1}$. This is a so far mostly unexplored regime in stellar mass and $\mathrm{SFR}$. ACE increases the number of unlensed galaxies at $z \sim 2$--$2.5$ for which both $\mathrm{CO}$ and metallicity information are available by a factor of approximately 3.  We compare the location of the galaxies in the stellar mass -- SFR parameterization to the star-forming main-sequence presented by \citet{Popesso2023}, which is based on a compilation of literature samples with SFRs predominantly obtained through SED fitting or from UV + IR emission. Most of the ACE galaxies lie within half a $\mathrm{dex}$ of the star-forming main sequence (all except for two) and have sSFRs between $10^{-9}$ and $10^{-8}\,\mathrm{yr}^{-1}$. 

The literature compilation, on the other hand, shows a significantly larger spread in $\mathrm{SFR}$s, resulting in a wide dispersion in $\mathrm{sSFR}$ and a substantial offset from the main sequence as well. The ACE sample shows a more normal distribution in most of the panels of Figure~\ref{fig:property_histograms}. This prominent difference in sample selection must be kept in mind when comparing the ACE galaxy properties to the literature compilation in all subsequent figures.

In Figure~\ref{fig:mstar_vs_sfr}, we show the placement of the ACE targets with respect to the main sequence of star-forming galaxies at $z \sim 2.3$ as derived by \citet{Popesso2023}. Most of the ACE galaxies are within $\sim 0.3\,\mathrm{dex}$ of the main sequence (see also Figure~\ref{fig:property_histograms}). A few of the galaxies, especially the two lowest-mass objects, are significantly elevated above the \citet{Popesso2023} main sequence by up to $\sim 1\,\mathrm{dex}$ for one of the targets. In contrast, the literature compilation shows a much larger scatter about the main sequence, with extremes extending up to an order of magnitude both above and below it.

\begin{figure*}
  \centering
    \includegraphics[width=\textwidth]{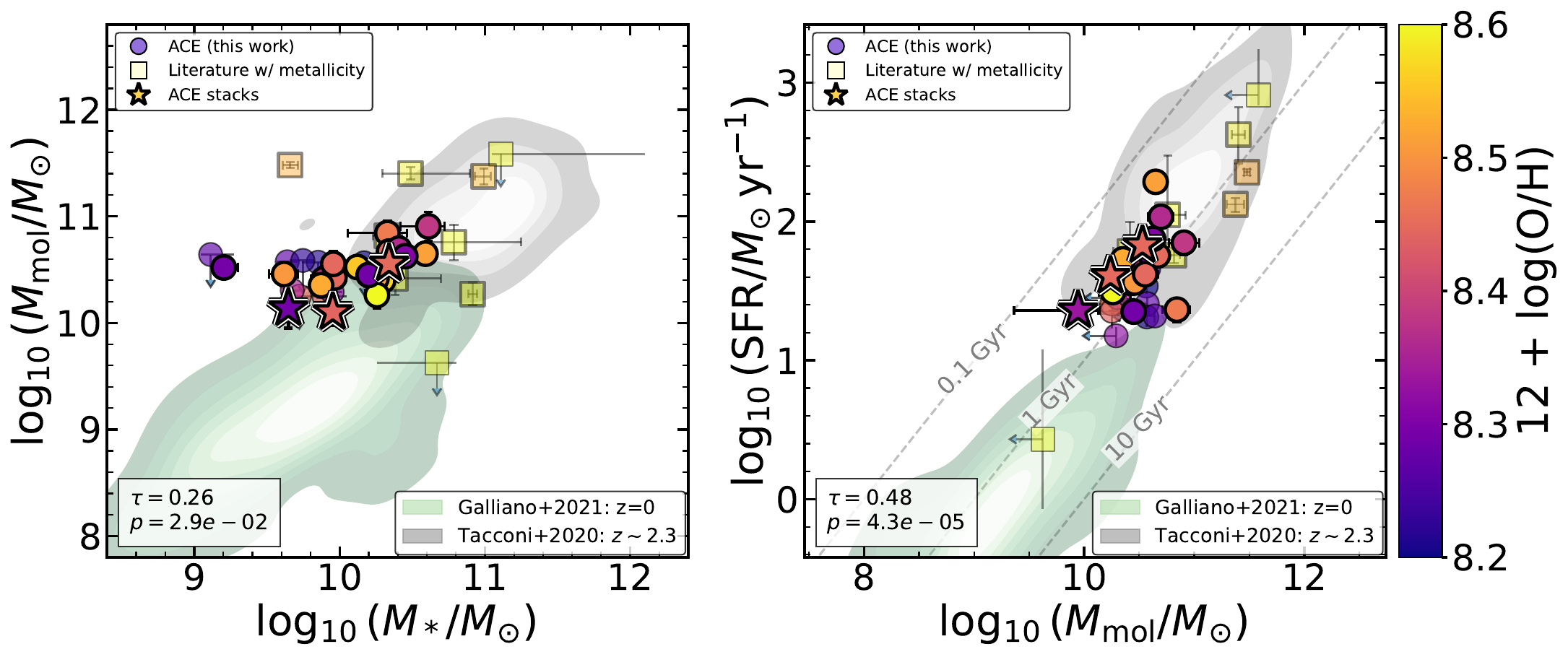}
  \caption{Molecular gas mass as a function of stellar mass (left panel) and star formation rate ($\mathrm{SFR}$) as a function of molecular gas mass (right panel), with lines of constant depletion time ($\tau_{\mathrm{dep}}$) overplotted. All points are colour-coded by gas-phase metallicity. In both panels, the ACE galaxies are marked as coloured circles, with stacked measurements depicted as stars. Galaxies from the literature compilation with available metallicity information are represented by squares. The compilation of galaxies from \citet{Tacconi2020} with $\mathrm{CO}$-based molecular gas masses is indicated by grey shaded contours, whereas the local $z=0$ comparison sample is shown with green shaded contours. In both panels, the Kendall rank correlation coefficient ($\tau$) and its accompanying $p$-value are listed in the bottom-left text box.
  \label{fig:Mstar_mmol_sfr}}
\end{figure*}

\subsection{Molecular gas scaling relations}
In the next three subsections, we show the scaling relations between molecular gas mass and other galaxy properties, focusing specifically on $M_{\mathrm{mol}}/M_{\star}$ and $t_{\mathrm{dep}}$. We show the individual and stacked measurements for the ACE sample. In all figures, we compare the ACE galaxies to the literature compilation with metallicity information, the \citet{Tacconi2020} compilation of galaxies with $\mathrm{CO}$-based molecular gas masses (assuming a fixed $\alpha_{\mathrm{CO}}$ conversion factor), and the $z=0$ comparison sample from \citet{Galliano2021}. The ACE galaxies, as well as the literature compilation with metallicity information, are all colour-coded according to their metallicity. Furthermore, we display the Kendall rank correlation coefficient in every figure to evaluate the degree of similarity between the two sets of ranks assigned to the same set of objects. We use the implementation presented in \citet{flury_kendall}, which is based on the two-variable correlation formalism of \citet{Akritas1996} that accounts for censoring (upper/lower limits). We evaluated the Kendall rank correlation based on the combined set of ACE galaxies and literature compilation with metallicity information. In all plots, $\tau$ denotes the degree of correlation, where 0 indicates no correlation, and increasingly positive (negative) values up to 1 (-1) represent a positive (negative) correlation. $p$ is the two-sided $p$-value, indicating the probability of obtaining the observed correlation purely by chance (with smaller $p$-values reflecting higher statistical significance). We present a full summary of the Kendall $\tau$ correlation coefficient and corresponding $p$-value for each relation plotted in this paper in Table~\ref{tab:kendall_censored_scaling}.

\begin{figure*}
  \centering
  \includegraphics[width=\textwidth]{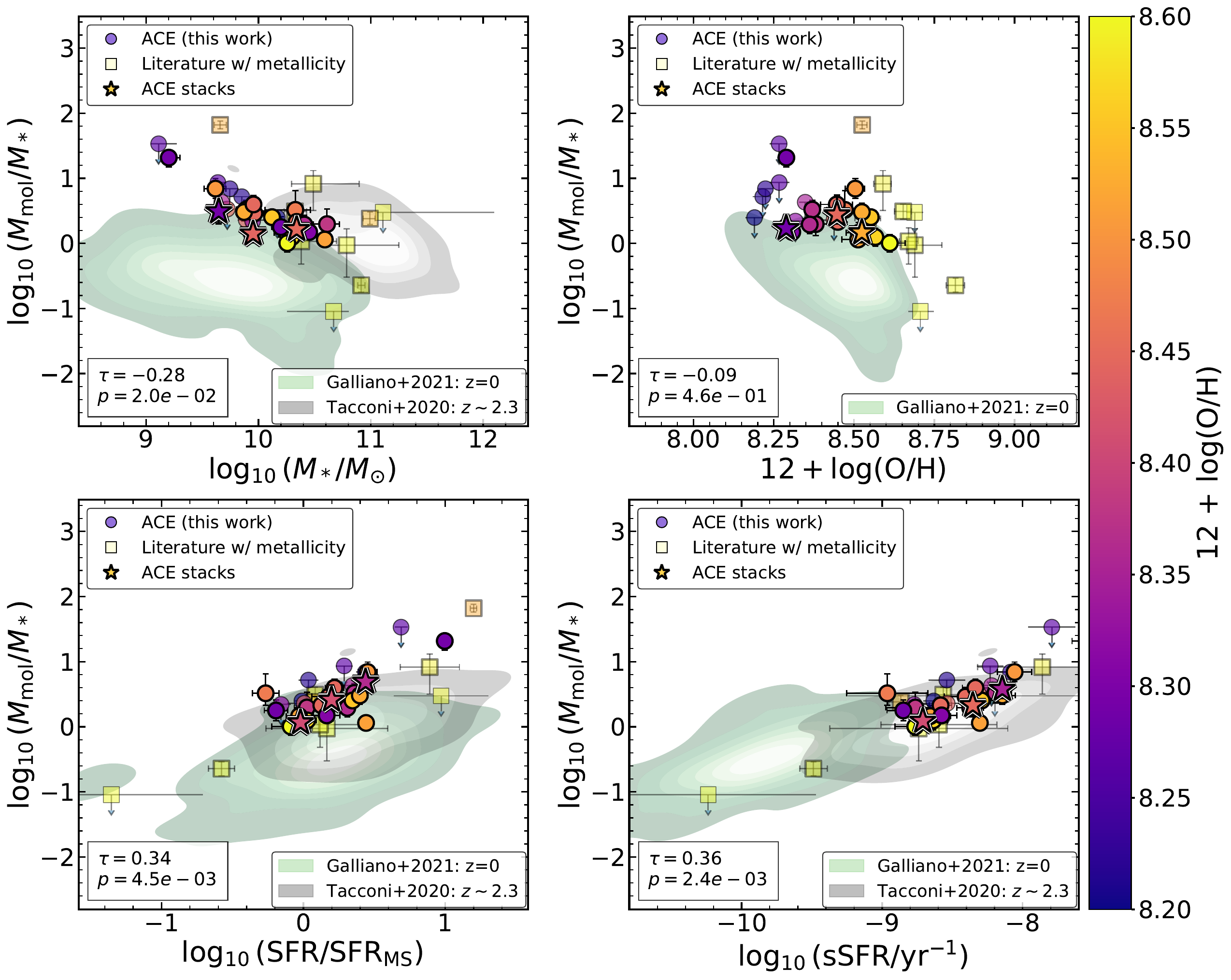}
  \caption{The molecular-to-stellar mass ratio, $M_{\mathrm{mol}}/M_{\star}$, of galaxies as a function of stellar mass (top-left panel), gas-phase metallicity (top-right panel), offset from the star-forming main sequence (bottom-left panel), and $\mathrm{sSFR}$ (bottom-right panel). All points are colour-coded by their gas-phase metallicity. The markers and shaded contours are identical to those in Figure~\ref{fig:Mstar_mmol_sfr}. In all panels the Kendall rank correlation coefficient ($\tau$) and its accompanying $p$-value are listed in the bottom-left text box. For the $z \sim 2$--$2.5$ ACE galaxies and the literature compilation with metallicity information, $M_{\mathrm{mol}}/M_{\star}$ exhibits a strong positive correlation with both the main-sequence offset and $\mathrm{sSFR}$, while showing a much weaker to negligible correlation with stellar mass and metallicity, respectively. 
  \label{fig:four_panel_mmol_over_mstar}}
\end{figure*}

\subsubsection{Molecular gas as a function of stellar mass and SFR}
We present the molecular gas mass of galaxies as a function of their stellar mass in the left panel of Figure~\ref{fig:Mstar_mmol_sfr}. We find that the ACE galaxies with a $\mathrm{CO}$ detection show a clear increase in their molecular gas mass with stellar mass of a dex over the stellar mass range from $10^{9.5}$ to $10^{10.5}\,\rm{M}_\odot$. The upper limits on molecular gas mass for targets without a robust $\mathrm{CO}$ detection are entirely consistent with this trend, though admittedly the stack of the lowest stellar mass bin has a molecular gas mass similar to the central bin. The latter may seem surprising at first glance, but it can be explained by a selection effect: our sample in the lowest stellar mass bin is preferentially biased toward galaxies occupying the upper envelope of, or lying entirely above, the star-forming main sequence (Figure~\ref{fig:mstar_vs_sfr}). Given the tight link between SFR and $M_{\mathrm{mol}}$ (see Figure~\ref{fig:Mstar_mmol_sfr}; right panel) this naturally drives a higher $M_{\mathrm{mol}}$ for a fixed stellar mass. Overall, the ACE galaxies form a natural lower-mass continuation of the $M_{\mathrm{mol}}$--$M_{\star}$ relation mapped out by the \citet{Tacconi2020} compilation. Crucially, the ACE sample probes galaxies down to an order of magnitude lower in stellar mass than probed before in unlensed galaxies at cosmic noon, demonstrating that this scaling relation persists down to at least $M_{\star} \sim 10^{9.5}\,\mathrm{M}_{\odot}$. The high-redshift literature compilation with metallicity information shows a significantly larger scatter in the $M_{\mathrm{mol}}$--$M_{\star}$ plane, reflecting its broader distribution in properties such as $\mathrm{SFR}$. We find a Kendall rank correlation coefficient of $\tau = 0.26$ with a $p$-value of $\sim 0.03$ for the combined ACE and literature sample, indicating a marginally significant positive correlation. The ACE galaxies possess molecular gas masses nearly a $\mathrm{dex}$ higher than their local $z = 0$ counterparts at a fixed stellar mass. We find no secondary dependence on metallicity within the ACE sample; that is, the residual scatter in the $M_{\mathrm{mol}}$--$M_{\star}$ plane is not explained by gas-phase metallicity. 

The right-hand panel of Figure~\ref{fig:Mstar_mmol_sfr} displays the $\mathrm{SFR}$ of galaxies as a function of their molecular gas mass, representing the integrated molecular Schmidt-Kennicutt relation \citep{Kennicutt1998, Daddi2010,Genzel2010}. The ACE galaxies define a remarkably tight correlation, where the $\mathrm{SFR}$ increases robustly with molecular gas mass. The ACE galaxies follow a somewhat super-linear relation between molecular gas mass and SFR, corresponding to an average molecular gas depletion time ($\tau_{\mathrm{dep}}$) just below $1\,\mathrm{Gyr}$. In contrast, the stacked measurements show a slightly sub-linear trend toward higher molecular gas masses. The differing slopes inferred from the detections and stacks likely reflect selection effects and the incomplete sampling of the star-forming main sequence across the full parameter space. When combined with literature measurements, however, the overall relation extends smoothly over a broad range of molecular gas masses and remains consistent with an approximately constant depletion time, in agreement with the compilation of \citet{Tacconi2020}. 

Because the ACE galaxies sample a lower stellar mass regime than previous $z \sim 2$--$2.5$ literature compilations, these data demonstrate that the integrated molecular Schmidt-Kennicutt relation extends to lower masses and lower star formation regimes. This tight scaling relation is reflected by a high Kendall correlation coefficient of $\tau = 0.48$ with a $p$-value of $4.3 \times 10^{-5}$, ruling out random chance as a driver. As with the relation between $M_{\mathrm{mol}}$ and $M_{\star}$, we find no strong secondary dependence on metallicity to account for the intrinsic scatter in the $M_{\mathrm{mol}}$--$\mathrm{SFR}$ relation. The super-linear trend between SFR and $M_{\rm mol}$ connects continuously down to the scales probed by the local $z = 0$ comparison sample, suggesting that the underlying relation is largely redshift-independent. It is important to note that we do not control for sample matching between the z$=$0 and cosmic noon populations. Therefore, the observed change in the average depletion time should not be interpreted as evidence for redshift evolution, as it may instead reflect differences in the sampled galaxy populations, such as their position relative to the main sequence or their specific star formation rates.

\begin{figure*}
  \centering
  \includegraphics[width=\textwidth]{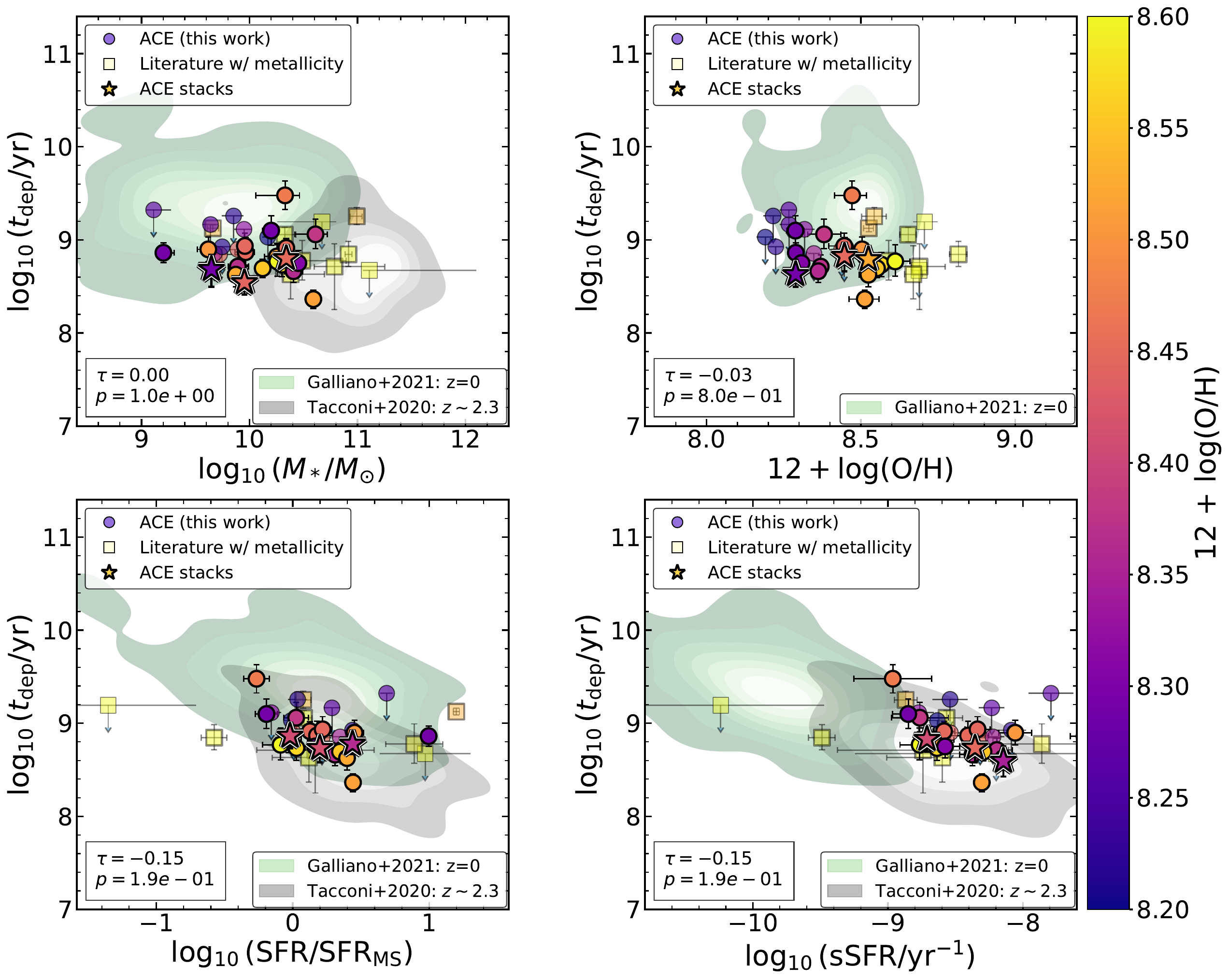}
  \caption{The molecular gas depletion time ($t_{\mathrm{dep}}$) of galaxies as a function of stellar mass (top-left panel), gas-phase metallicity (top-right panel), offset from the star-forming main sequence (bottom-left panel), and $\mathrm{sSFR}$ (bottom-right panel). All points are colour-coded by their gas-phase metallicity. The markers and shaded contours are identical to those in Figure~\ref{fig:Mstar_mmol_sfr}. In all panels the Kendall rank correlation coefficient ($\tau$) and its accompanying $p$-value are listed in the bottom-left text box. For the $z \sim 2$--$2.5$ ACE galaxies and the literature compilation, $t_{\mathrm{dep}}$ exhibits a very weak correlation with both the main-sequence offset and $\mathrm{sSFR}$, and no significant correlation with either stellar mass or metallicity. 
  \label{fig:four_panel_tdep}}
\end{figure*}
\subsubsection{The molecular-to-stellar mass ratio}
In Figure~\ref{fig:four_panel_mmol_over_mstar}, we show the molecular-to-stellar mass ratio ($M_{\mathrm{mol}}/M_{\star}$) of galaxies as a function of their stellar mass, metallicity, offset from the star-forming main sequence ($\mathrm{SFR}/\mathrm{SFR}_{\mathrm{MS}}$), and $\mathrm{sSFR}$, respectively. We find a weakly declining trend between $M_{\mathrm{mol}}/M_{\star}$ and stellar mass, with a Kendall rank correlation coefficient of $\tau = -0.28$. This downward trend is also apparent in the \citet{Tacconi2020} and local $z=0$ compilations, though it is less pronounced. Similarly, the stacked binned measurements of the ACE sample show a shallower trend than the individual galaxies. We find no secondary dependence on metallicity driving the scatter within the $M_{\mathrm{mol}}/M_{\star}$--$M_{\star}$ plane. We find that at fixed stellar mass, $M_{\mathrm{mol}}/M_{\star}$ increases with approximately an order of magnitude between $z=0$ to $z=2-2.5$, across the full stellar mass range probed. The two lowest-mass ACE galaxies have the highest molecular gas fractions ($M_{\mathrm{mol}}/M_{\star}$), potentially suggesting a steeper increase of $M_{\mathrm{mol}}/M_{\star}$ toward lower stellar masses than inferred from the remainder of the sample. However, it is important to note that both galaxies lie significantly above the star-forming main sequence and may therefore not be representative of the typical galaxy population at these masses.

We observe a weak correlation between $M_{\mathrm{mol}}/M_{\star}$ and gas-phase metallicity for the ACE galaxies, the cosmic noon literature compilation with metallicity information, and the stacked data, confirmed by a Kendall coefficient of $\tau = -0.09$ with a $p$-value of $0.46$. This lack of a strong correlation is somewhat surprising, given the prominent decline in $M_{\mathrm{mol}}/M_{\star}$ as a function of increasing metallicity seen in the local $z=0$ comparison sample. The differences may be driven by selection effects, both for the $z=0$ sample and the cosmic noon galaxies. Indeed, the number of low-metallicity galaxies in \citet{Galliano2021} with CO measurement counterparts is very limited, restricted to the CO brightest (and thus most molecular gas rich) galaxies. Furthermore, we are not comparing samples that are complete in the multi-dimensional parameter space of stellar mass, SFR and metallicity at their respective redshifts. We do clearly find that the cosmic noon galaxies have higher $M_{\mathrm{mol}}/M_{\star}$ at fixed metallicity than the $z=0$ compilation, with a typical difference of approximately 0.5 dex. This is in agreement with predictions by theoretical models \citep[e.g.,][]{Popping2015, Lagos2015, Dave2020}.

The bottom-left and bottom-right panels of Figure~\ref{fig:four_panel_mmol_over_mstar} demonstrate a clear positive correlation between $M_{\mathrm{mol}}/M_{\star}$ and both the main-sequence offset and $\mathrm{sSFR}$, respectively, with Kendall values of $\tau = 0.35$ ($p = 0.0045$) and $\tau = 0.30$ ($p = 0.0024$). The presence of a star formation term on the $x$-axis of both scaling relations underscores the tight physical coupling between $\mathrm{SFR}$ and molecular gas mass (as shown in the right panel of Figure~\ref{fig:Mstar_mmol_sfr}). This robust positive correlation is also clearly visible in the stacked data points. We find no obvious secondary dependence on metallicity in either of these trends. 

We find a distinct redshift evolution in $M_{\mathrm{mol}}/M_{\star}$ of approximately one dex from $z=0$ to $z=2-2.5$ at a fixed offset from the main sequence. The positive correlation between $M_{\mathrm{mol}}/M_{\star}$ and $\mathrm{sSFR}$ on the other hand is redshift independent. The latter is a natural consequence of the tight relation between $\mathrm{SFR}$ and $M_{\mathrm{mol}}$, since $M_{\mathrm{mol}}/M_{\star}$ and $\mathrm{sSFR}$ effectively share $M_{\star}$ as a common denominator. The redshift evolution in $M_{\mathrm{mol}}/M_{\star}$ at fixed stellar mass can thus also be explained as an evolution in sSFR.

\subsubsection{The molecular gas depletion time}
We show the molecular gas depletion time of galaxies as a function of their stellar mass, metallicity, offset from the star-forming main sequence, and $\mathrm{sSFR}$ in Figure~\ref{fig:four_panel_tdep}. We find little to no trend between the depletion time of the galaxies and either their stellar mass or metallicity, yielding Kendall rank correlation coefficients of $\tau = 0.00$ and $\tau = -0.03$, respectively. This lack of correlation is also apparent from the stacked binned measurements. The depletion times of the ACE galaxies are centered around $1\,\mathrm{Gyr}$ (ranging from $\sim 300\,\mathrm{Myr}$ up to $\sim 3\,\mathrm{Gyr}$), in good agreement with depletion times previously found for more massive galaxies at these epochs (see the literature compilation with metallicity, as well as, e.g., \citealt{Aravena2019, Tacconi2020}). On average, the ACE galaxies have depletion times that are somewhat shorter than those found for the local $z=0$ comparison sample.

There is a slightly more pronounced downward trend between depletion time and both the offset from the main sequence and $\mathrm{sSFR}$, with a Kendall coefficient of $\tau = -0.15$ ($p = 0.19$) for both parameters. These trends are in good agreement with the \citet{Tacconi2020} compilation and are also clearly reflected in the stacked bins. This indicates that galaxies become increasingly efficient at forming stars as they move from below to above the star-forming main sequence. We find that at a fixed offset from the main sequence, the ACE galaxies have depletion times that are systematically shorter than those of $z=0$ galaxies (with typical values of $\sim 1\,\mathrm{Gyr}$ versus $\sim 3\,\mathrm{Gyr}$). Similar to the behaviour observed for $M_{\mathrm{mol}}/M_{\star}$, there appears to be an underlying relation between $\mathrm{sSFR}$ and $t_{\mathrm{dep}}$ that holds across cosmic time, where galaxies with higher $\mathrm{sSFR}$ have  shorter gas depletion times.

\begin{figure*}
  \centering
  \includegraphics[width=0.8\textwidth]{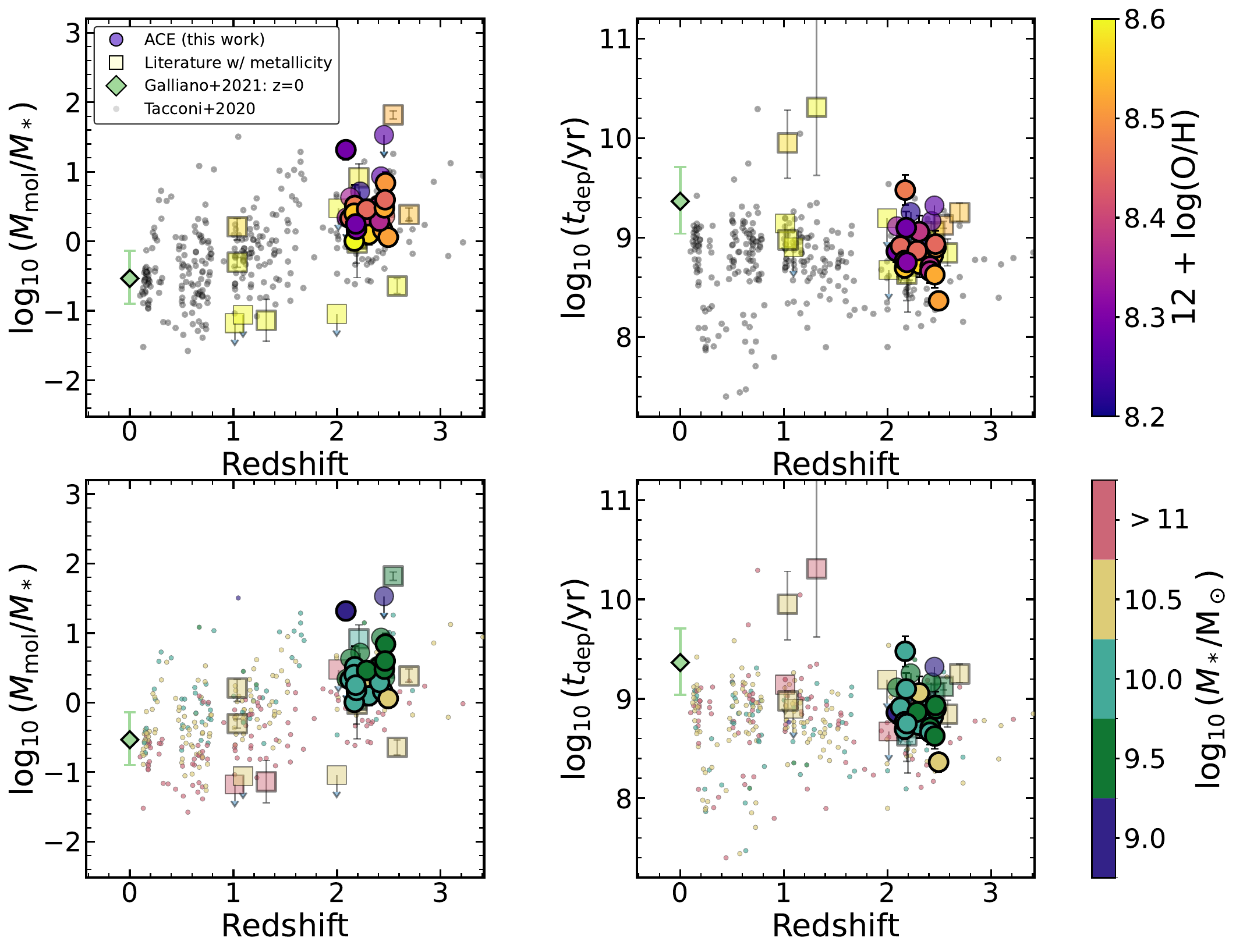}
  \caption{The molecular-to-stellar mass ratio ($M_{\mathrm{mol}}/M_{\star}$; left panels) and molecular gas depletion time ($t_{\mathrm{dep}}$; right panels) of galaxies as a function of redshift, colour-coded by their gas-phase metallicity (top row) and stellar mass (bottom row). Circles mark the ACE galaxies, while squares indicate galaxies from the literature compilation with available metallicity information (note that here we also include targets at $z < 2$). Small circles represent galaxies from the \citet{Tacconi2020} literature compilation of sources with $\mathrm{CO}$-based molecular gas masses, and the diamond marks the median of the local $z=0$ compilation from \citet{Galliano2021}. On average, we find that $M_{\mathrm{mol}}/M_{\star}$ decreases by nearly an order of magnitude from $z \sim 2$--$2.5$ down to $z=0$, whereas $t_{\mathrm{dep}}$ remains largely constant over cosmic time, showing at most a slight increase of a factor of $\sim 2$--$3$. Furthermore, a significant scatter of roughly an order of magnitude is present in both $M_{\mathrm{mol}}/M_{\star}$ and $t_{\mathrm{dep}}$ at any given redshift. There is no clear secondary dependence on metallicity within the scaling relations shown that drives this observed scatter, but we do find that lower mass galaxies on average have higher $M_{\mathrm{mol}}/M_{\star}$.
  \label{fig:redshift_two_panel}}
\end{figure*}
\subsection{Redshift evolution}
We show the molecular-to-stellar mass ratio and molecular gas depletion time of galaxies as a function of redshift in Figure~\ref{fig:redshift_two_panel}, colour-coded by their gas-phase metallicity and stellar mass. We find that $M_{\mathrm{mol}}/M_{\star}$ increases by approximately an order of magnitude from $z = 0$ to $z \sim 2$--$2.5$. Conversely, the depletion time decreases slightly with redshift, by roughly a factor of three over the same cosmic timeframe. These results are in excellent agreement with previous evolutionary studies \citep[e.g.,][]{Aravena2019, Tacconi2020}. 

Notably, this evolutionary trend has not previously been investigated for unlensed galaxies with stellar masses of $M_{\star} < 10^{10.5}\,\mathrm{M}_{\odot}$ and metallicities extending well below solar. We show for the first time that galaxies in this low-mass, low-metallicity regime follow the same redshift evolution in $M_{\mathrm{mol}}/M_{\star}$ between $z \sim 2.3$ and $z = 0$ as observed in more massive systems, while maintaining systematically higher molecular gas fractions at fixed redshift (Figure~\ref{fig:four_panel_mmol_over_mstar}).

\section{Discussion}
\label{sec:discussion}
\subsection{What drives star formation in galaxies?}
A central objective in studying galaxy assembly is understanding the fundamental drivers of star formation. Is an elevated SFR primarily driven by an increased supply of available molecular gas mass—the raw fuel for star formation—or by a fundamental shift in star formation efficiency? Furthermore, does this efficiency evolve as a function of stellar mass, gas-phase metallicity, different galaxy properties or time? In this paper, we present the molecular gas properties of 26 typical star-forming galaxies at $z \sim 2$--$2.5$, observed with ALMA as part of the ACE large program. The defining feature of this sample is that it probes unlensed galaxies with stellar masses nearly an order of magnitude lower, and metallicities $\sim 0.4\,\mathrm{dex}$ lower, than previously achieved at cosmic noon by surveys such as PHIBSS \citep{Tacconi2018} and ASPECS \citep{Aravena2019}. This opens up a new parameter space for investigating molecular gas and its link to the stellar buildup in typical galaxies at cosmic noon.

Our results support the view that molecular gas availability is the primary regulator of star formation. This is clearly demonstrated by the tight correlation between $M_{\mathrm{mol}}$ and SFR (Figure~\ref{fig:Mstar_mmol_sfr}; right panel), which is the most statistically significant relation identified in this work and underscores the fundamental role of molecular gas availability in driving star formation. Moreover, the smooth continuity of this relation between the ACE galaxies and $z=0$ comparison sample suggests that its overall form is largely independent of redshift. Although the observed differences in average depletion times between cosmic noon and $z=0$ could indicate evolution, they should be interpreted with caution, as the low- and high-redshift samples are not matched in their galaxy properties. Any apparent variation may therefore reflect differing distributions in specific star formation rate or position relative to the star-forming main sequence \citep{Saintonge2022}, rather than genuine redshift evolution. 

Consistent with the interpretation that variations in depletion time are primarily driven by galaxy properties rather than redshift, Figures~\ref{fig:four_panel_mmol_over_mstar} and \ref{fig:four_panel_tdep} show that both the molecular-to-stellar mass ratio and molecular gas depletion time correlate most strongly with $\mathrm{sSFR}$ and offset from the star-forming main sequence. These relations remain continuous across the combined $z \sim 0$ and $z \sim 2$ samples, further strengthening that $\mathrm{sSFR}$ is closely linked to molecular gas fraction and depletion time across cosmic time. The prominent correlation with $\mathrm{sSFR}$ and main-sequence offset, characterized by Kendall rank coefficients of $\tau \sim 0.35$ for $M_{\mathrm{mol}}/M_{\star}$ and $\tau = -0.15$ for $t_{\mathrm{dep}}$, reveals two key physical insights. First, at fixed stellar mass, galaxies with elevated $\mathrm{sSFR}$ (above the main sequence) possess larger molecular gas reservoirs, demonstrating that gas availability is the primary driver of enhanced star formation activity. Second, these galaxies also deplete their molecular gas more rapidly, implying higher star formation efficiencies than systems with lower $\mathrm{sSFR}$ (below the main sequence). These trends remain coherent across a wide stellar mass range of $10^{9}$--$10^{11}\,\mathrm{M}_{\odot}$ and a gas-phase metallicity range of $12 + \log(\mathrm{O/H}) = 8.2$--$8.8$, indicating that variations in both gas content and star formation efficiency govern a galaxy's position relative to the star-forming main sequence. However, the physical mechanisms responsible for the observed efficiency variations remain uncertain.


We find no significant correlation between molecular gas depletion time and gas-phase metallicity (top-right panel of Figure~\ref{fig:four_panel_tdep}). One might expect that the reduced shielding against ionizing radiation in metal-poor environments would alter molecular cloud structures, rendering them less efficient at forming stars. Complementary work by \citet{Geesink2026} demonstrates that the dust-to-gas ratio increases with metallicity, and \citet{Popescu2026} find that dust temperatures systematically increase toward lower metallicities. Nevertheless, these prominent variations in interstellar medium  properties do not appear to impact the conversion efficiency of molecular gas into stars across the entire metallicity regime covered by ACE. This unexpected invariance could be indicative of a relatively constant dense gas fraction (i.e., the dense molecular gas cores where star formation actually occurs), though confirming this scenario will require future observations of high-density gas tracers. We emphasize that this work focuses exclusively on the molecular gas budget. Because theoretical models predict a strong correlation between the fraction of cold gas in the molecular phase ($f_{\mathrm{H}_2}$) and gas-phase metallicity \citep{Krumholz2008, Gnedin2010}, one might expect a significantly more pronounced trend to emerge when considering the total cold-gas depletion time (including atomic hydrogen) of galaxies as a function of their metallicity. A caveat of our analysis is furthermore that the metallicities considered are derived from \hii region emission lines and therefore trace the ionized gas phase rather than the molecular gas itself. Implicitly, this assumes that the metallicity of the ionized gas is representative of that of the molecular gas. Any systematic differences between these phases could weaken intrinsic trends and may partly account for the lack of a significant correlation with metallicity found in our analysis.

Future analyses of spatially resolved gas, stellar mass and star-formation distributions of the ACE galaxies will be crucial for establishing how $M_{\mathrm{mol}}/M_{\star}$ and $t_{\mathrm{dep}}$ are connected to the underlying dynamical state and stability of galactic discs. Such observations will clarify the physical drivers behind the increased star formation efficiency observed as a function of sSFR, extending studies of more massive galaxies \citep[see the review by][]{Tacconi2020}.

\begin{figure}
  \centering
  \includegraphics[width=\columnwidth]{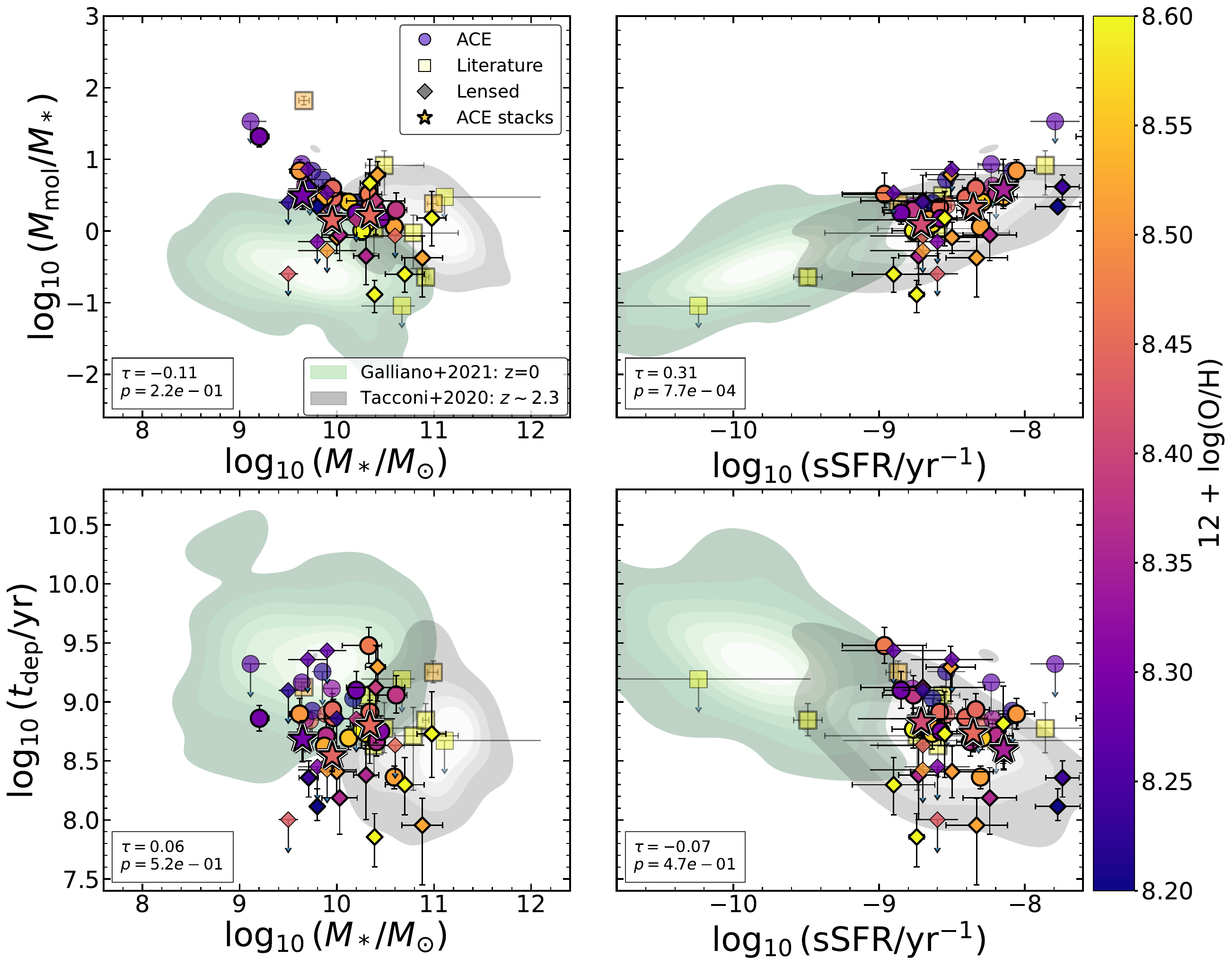}
  \caption{The molecular-to-stellar mass ratio ($M_{\mathrm{mol}}/M_{\star}$; top row) and molecular gas depletion time ($t_{\mathrm{dep}}$; bottom row) as a function of stellar mass and  specific star formation rate (sSFR). The data points are colour-coded by gas-phase metallicity. Circles denote the ACE galaxies, stars the ACE stacks and squares represent galaxies from the literature compilation with available metallicity measurements. Small grey diamonds indicate the strongly lensed galaxies from \citet{Catan2024}, \citet{Saintonge2013} and lensed galaxies A68-HLS115 \citep{DessaugesZavadsky2015,Girard2018}, SDSS J0901+1814 \citep{Sharon2019}, MACS J0451+0006 (MACS J0451 Arc) \citep{DessaugesZavadsky2015,Curti2020}, and RCSGA 032727-132609 \citep{Rigby2011, Wuyts2014, GonzalezLopez2017}.  For all lensed galaxies we recomputed the molecular gas masses using the metallicity-dependent CO-to-H$_2$ conversion factor of \citet{Accurso2017}, but the SFR and stellar mass are as listed in the respective papers based on different methodologies from the ACE sample. The grey and green shaded regions show the distributions of galaxies from the compilations of \citet{Tacconi2020} and \citet{Galliano2021}, respectively. In all panels the Kendall rank correlation coefficient ($\tau$) and its accompanying $p$-value are listed in the bottom-left text box. 
  \label{fig:scaling_compare_lenses}}
\end{figure}

\subsection{Placing ACE in context: a comparison to lensed galaxies}
This work presents some of the first constraints on the molecular gas content and depletion times of unlensed galaxies at cosmic noon spanning stellar masses of $M_{\star} < 10^{10.5}\,\mathrm{M}_{\odot}$ and gas-phase metallicities of $12 + \log(\mathrm{O/H}) < 8.6$. As discussed above, the scaling relations established for more massive and metal-rich galaxies appear to remain valid within this newly explored parameter space. However, it is important to account for the incompleteness of the ACE sample. As shown in Figure~\ref{fig:mstar_vs_sfr} and discussed by \citet{Shivaei2026}, the ACE selection preferentially targets galaxies located on or slightly above the star-forming main sequence, particularly at low stellar masses and metallicities, and therefore misses galaxies in the lower envelope of the main sequence. 

To place the ACE results into a broader context, we can compare them to studies of low-mass galaxies observed through gravitational lensing \citep[e.g.,][]{Saintonge2013, DessaugesZavadsky2015, Sharon2019, Catan2024}. The molecular gas fractions and depletion times measured for the ACE galaxies are generally consistent with those reported for lensed galaxy samples, suggesting that the gas scaling relations observed here extend into the low-mass and low-metallicity regime probed by ACE.

In Figure~\ref{fig:scaling_compare_lenses}, we compare a compilation of lensed sources with metallicity and low-J CO information at redshifts $z\sim 1.5 - 3$ to the ACE galaxies and literature compilation with metallicity information. This compilation consists of all targets presented in \citet{Saintonge2013} and \citet{Catan2024}, as well as A68-HLS115 \citep{DessaugesZavadsky2015,Girard2018}, SDSS J0901+1814 \citep{Sharon2019}, MACS J0451+0006 (MACS J0451 Arc) \citep{DessaugesZavadsky2015,Curti2020}, and RCSGA 032727-132609 \citep{Rigby2011, Wuyts2014, GonzalezLopez2017}. When making this comparison, it is important to note that the stellar masses and SFRs of the lensed galaxies were derived using methodologies different from those adopted in this work. We recomputed the molecular masses presented in the respective papers using the metallicity-dependent calibration of \citet{Accurso2017} for consistency with the methodology employed throughout the ACE survey. \footnote{\citet{Catan2024} infer metallicities from stellar population modelling following \citet{Solimano2022}, adopting the simplifying assumption that the stellar and gas-phase metallicities are identical. No direct gas-phase metallicity measurements are available for these galaxies.}


We find that the lensed galaxies broadly overlap with the ACE sample and follow the same general trends in molecular gas fraction and depletion time. At fixed stellar mass, the lensed galaxies are largely consistent with the ACE population and do not significantly broaden the observed distribution. However, when considering sSFR, the upper limits of the lensed sample push toward lower molecular gas fractions and shorter depletion times than the limits of the ACE sample. In Figure~\ref{fig:scaling_compare_lenses}, we additionally report the Kendall rank correlation coefficients for the combined ACE, literature, and lensed galaxy samples. We find that both the resulting Kendall $\tau$ values and the associated two-sided $p$-values remain qualitatively unchanged compared to those listed in Table~\ref{tab:kendall_censored_scaling} for the ACE and metallicity-selected literature sample alone. This indicates that the inclusion of the lensed galaxies strengthens the explored parameter space without significantly altering the overall correlations.

This comparison between lensed and unmagnified galaxies highlights the importance of pursuing completeness across the full multi-dimensional parameter space of stellar mass, metallicity, and star-formation activity. Although ACE has opened a new window on the low-mass and low-metallicity galaxy population at cosmic noon, future surveys will require both greater sensitivity and more complete sampling of the star-forming main sequence to fully characterize the diversity of molecular gas properties in this regime.

\subsection{Implications for galaxy formation theory}
The observations presented here are of particular interest to cosmological galaxy formation simulations. These provide predictions for the (molecular) gas content and properties of large populations of galaxies across a wide stellar mass range, and robust observational constraints across the predicted mass range are thus essential for testing the physical processes that regulate the conversion of atomic gas into molecular gas and ultimately into stars (see \citealt{Dave2020} for a comprehensive comparison of the cold gas properties of a number of cosmological simulations). The link to gas-phase metallicity is especially relevant, as many simulations adopt a metallicity dependent recipe to calculate the fraction of cold gas that is in a molecular form. 

We find a weakly declining trend between molecular gas fraction, $M_{\rm mol}/M_\star$, and stellar mass over the range probed by our data (Figure~\ref{fig:four_panel_mmol_over_mstar}, top-left panel), in broad agreement with the behaviour predicted by several galaxy formation models \citep[e.g.,][]{Lagos2011,Popping2015,Dave2020}.  We further find that galaxies in our sample follow the integrated molecular Schmidt--Kennicutt relation (right panel of Figure~\ref{fig:Mstar_mmol_sfr}), linking molecular gas to SFR. This supports the use of the (resolved variant of the) relation as a star formation prescription in galaxy evolution models. Indeed, many semi-analytic and cosmological models assume by construction that star formation proceeds from the molecular gas phase following a variant of the Schmidt-Kennicutt relation \citep{Somerville2015}. The extension of the relation between molecular gas mass and SFR at cosmic noon down to lower stellar masses and metallicities, provides empirical support for this widely adopted framework. Note however, that more recent cosmological scale simulations have begun to explicitly model the formation of molecular hydrogen from first-principles \citep{Ploeckinger2020,RagoneFigueroa2024, Schaye2026} and decouple the SFR from the molecular gas content \citep{Schaye2026}. This development enables the relation between molecular gas content and star formation to emerge as a genuine prediction of galaxy formation simulations, rather than being imposed by construction (see, e.g., \citealt{Lagos2026}). The observational results presented here thus offer important constraints for assessing the ability of these models to reproduce the physical processes governing galaxy evolution.
 
We find no strong dependence of the molecular gas depletion time on gas-phase metallicity (Figure~\ref{fig:four_panel_tdep}, top-right panel). This result differs from the trends predicted by the  COLIBRE simulations \citep{Lagos2026}, who find systematically longer molecular gas depletion times at higher metallicities at resolved scales. At the same time, our data reveal a clear dependence of depletion time on specific star formation rate (Figure~\ref{fig:four_panel_tdep}, bottom-right panel), in qualitative agreement with the COLIBRE predictions \citet{Lagos2026}. The apparent lack of a strong metallicity trend in the observations should nevertheless be interpreted with caution. First of all, the current work presents integrated properties, whereas \citet{Lagos2026} show resolved properties. Second, selection effects may significantly influence the observed parameter space and potentially obscure intrinsic correlations. A more detailed comparison between observations and theory, including a careful treatment of survey selection functions and detection biases, is therefore required.

\begin{figure}
  \centering
    \centering
    \includegraphics[width=\columnwidth]{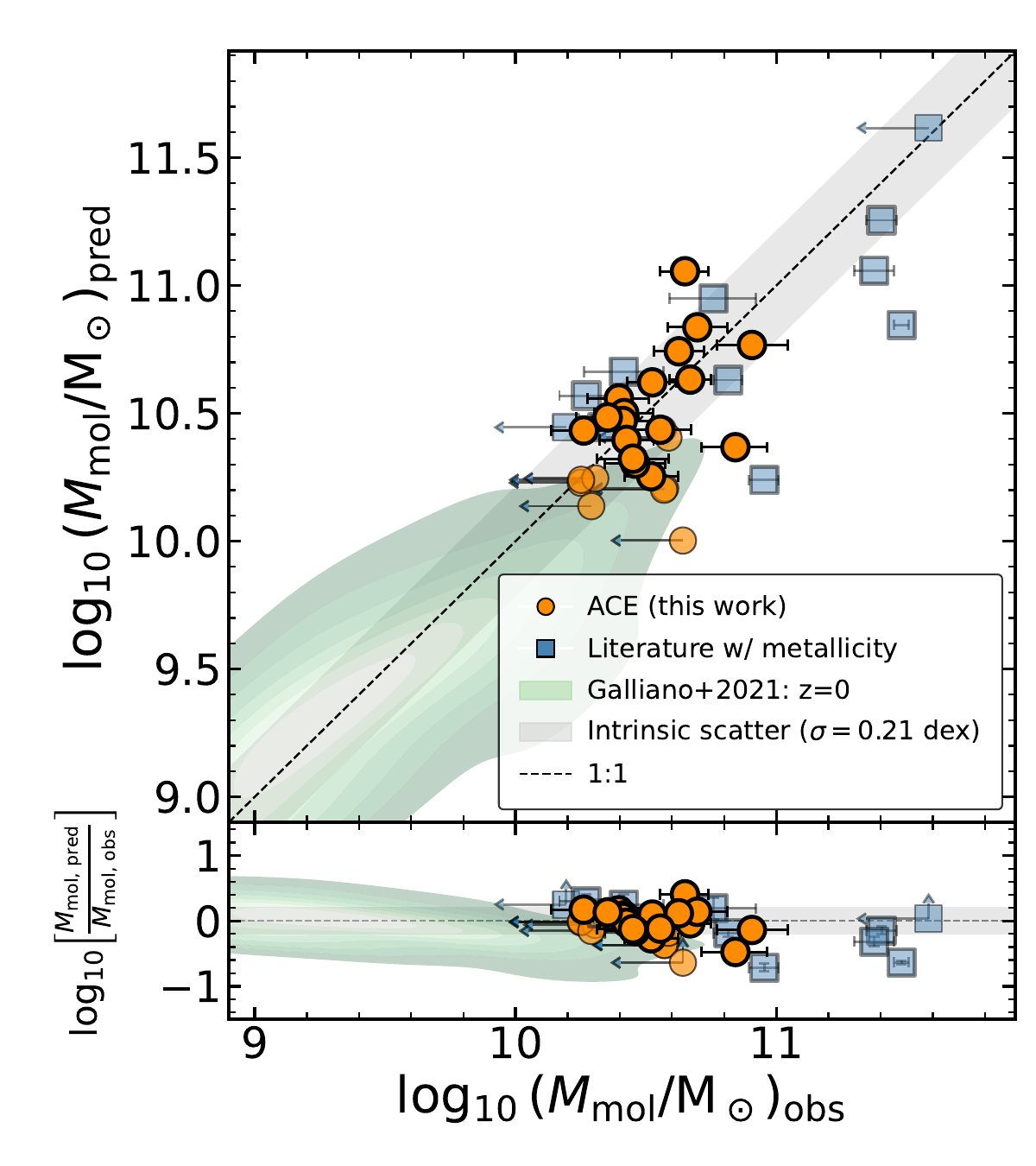}
  \caption{Predicted $(M_{\mathrm{mol}}$, obtained from Equation~\ref{eq:fitting_equation} as a function of the observed $(M_{\mathrm{mol}}$, for the ACE sample, literature compilation with available metallicity measurements and the $z=0$ sample (top panel). The dashed line denotes the one-to-one relation, while the shaded region indicates the intrinsic scatter of the fit. The bottom panel shows the ratio $(M_{\mathrm{mol}})_{\mathrm{pred}}/(M_{\mathrm{mol}})_{\mathrm{obs}}$ as a function of the observed $(M_{\mathrm{mol}}$. Equation~\ref{eq:fitting_equation} successfully reproduces the observed molecular gas mass across multiple orders of magnitude in $M_{\mathrm{mol}}$ and over the full redshift range probed by the samples.
  \label{fig:fitting_equation}}
\end{figure}

\subsection{Extending cold gas prediction relations to lower masses}
The presented observations of unlensed cosmic-noon galaxies spanning a wider range of stellar masses and metallicities than published thus far provide an ideal opportunity to reassess molecular-gas scaling relations and derive updated prescriptions for predicting $M_{\rm mol}/M_\star$ across an expanded parameter space. This builds directly upon previous scaling frameworks, such as those presented by \citet{Liu2019}, \citet{Tacconi2020}, and \citet{Sanders2023}. A key advantage of the galaxies presented in this work is the baseline homogeneity. As all $\mathrm{H}_2$ masses in this work are derived using a uniform methodology (also for the derivation of gas-phase metallicities), our sample avoids the potential systematic effects introduced by mixing different gas tracers, such as the scaling relations of \citet{Tacconi2020} which rely on a combined compilation of both $\mathrm{CO}$- and dust-based molecular gas masses.

In Figure~\ref{fig:Mstar_mmol_sfr} we found a strong trend between $M_{\mathrm{mol}}$ and SFR over 2 orders of magnitude SFR and cosmic time. We therefore adopt a functional form linking $M_{\mathrm{mol}}$ and SFR to each other. In Figure~\ref{fig:four_panel_tdep} we found a weak dependence between sSFR and molecular gas depletion time and we therefore add sSFR as a second parameter. We fit $\log_{10}(M_{\mathrm{mol}})$ using a Bayesian linear regression where the intercept, slopes, and intrinsic scatter ($\sigma$) are treated as free parameters. The posterior distributions are explored using MCMC sampling. For secure detections, we adopt a Gaussian likelihood that propagates asymmetric uncertainties in $\log_{10}(M_{\mathrm{mol}}/M_{\star})$. $\mathrm{CO}$ non-detections are incorporated as upper limits via a censored likelihood, ensuring that the true molecular gas masses are properly bounded. We exclude any literature galaxies with SFRs that classify them as quenched, defined here as lying more than $0.6\,\mathrm{dex}$ below the \citet{Popesso2023} main sequence. Our best-fit relation is given by:
\begin{equation}
\begin{split}
\log_{10}\!\left(M_{\mathrm{mol}}/\mathrm{M}_\odot\right) &= (0.94 \pm 0.03)\,\log_{10}\!\left(\mathrm{SFR}/\mathrm{M}_\odot\,\mathrm{yr}^{-1}\right)\\ 
&\quad- (0.28 \pm 0.04)\,\log_{10}\!\left(\mathrm{sSFR}/10^{-9}\,\mathrm{yr}^{-1}\right) \\
&\quad + (9.09 \pm 0.04) \\
\end{split}
\label{eq:fitting_equation}
\end{equation}

with an intrinsic scatter of $\sigma = 0.20 \pm 0.02$ dex. Figure~\ref{fig:fitting_equation} shows the predicted molecular gas mass from this best-fit relation against the observed values, along with the corresponding residuals. We find good agreement across the entire probed range of $M_{\mathrm{mol}}$. 

It is worth noting that metallicity is not explicitly included as an independent parameter in our fitting equation. Instead, it implicitly enters the relation through the combination of SFR and sSFR (which together connect directly to $\mathrm{M}_*$). Because $\mathrm{SFR}$ and stellar mass are two of the primary parameters defining the Fundamental Metallicity Relation \citep[FMR; e.g.,][]{Mannucci2010, LaraLopez2010}, the metallicity information is effectively encoded within these free parameters. A more detailed exploration of this, specifically focusing on the interplay between gas mass, metallicity, and stellar mass as a gas-based alternative to the FMR, is presented in \citet{Langan2026}.

\section{Summary}
\label{sec:summary}

In this work, we presented a detailed study of molecular gas scaling relations in a sample of low-mass, low-metallicity star-forming galaxies at $z \sim 2$--$2.5$, observed as part of the ALMA Chemical Evolution (ACE) survey. By targeting CO J$=$3--2 emission in unlensed galaxies with stellar masses down to $M_\star \sim 10^{9}\,\mathrm{M}_\odot$ and metallicities significantly below those of previous samples ($8.2 < 12 + \log(\mathrm{O/H}) < 8.6$), we substantially extend the explored parameter space of molecular gas studies at cosmic noon. Combining these observations with a carefully homogenized literature compilation and consistent metallicity-dependent $\alpha_{\mathrm{CO}}$ estimates, we investigated how molecular gas content, gas fractions, and depletion times scale with global galaxy properties. The main findings of this work can be summarized as follows:

\begin{itemize}
    \item \textbf{Molecular gas scaling relations at cosmic noon extend to $M_{\star} \sim 10^{9.5}\,\mathrm{M}_\odot$:} The ACE galaxies follow the established $M_{\mathrm{mol}}$--$M_{\star}$ and $M_{\mathrm{mol}}$--SFR relations, extending them down to at least $M_{\star} \sim 10^{9.5}\,\mathrm{M}_\odot$ and metallicities of $12 + \log{(O/H}) \sim 8.2$. This demonstrates that the fundamental scaling relations identified in massive galaxies remain valid in the low-mass, low-metallicity regime at cosmic noon.
    
    \item \textbf{Tight coupling between molecular gas and star formation:} We recover a strong, slightly super-linear correlation between molecular gas mass and SFR. This relation has the tightest correlation in our analysis ($\tau = 0.48$), indicating that the availability of molecular gas is the primary regulator of star formation activity.
    
    \item \textbf{Offset from the main-sequence primarily driven by availability of molecular gas reservoir:} The molecular-to-stellar mass ratio ($M_{\mathrm{mol}}/M_{\star}$) correlates strongly with  $\mathrm{sSFR}$ and offset from the star-forming main sequence at fixed redshift, although showing only weak or negligible dependence on stellar mass and metallicity. This highlights that variations in the available gas reservoir are the dominant driver of elevated star formation rates at fixed stellar mass.
    
    \item \textbf{Weak dependence of depletion time on galaxy properties:} The molecular gas depletion time ($t_{\mathrm{dep}}$) shows little to no correlation with stellar mass or metallicity, and only a weak dependence on main-sequence offset and $\mathrm{sSFR}$. Typical depletion times are $\sim 1\,\mathrm{Gyr}$, consistent with previous studies at similar redshifts.
    
    \item \textbf{No strong metallicity dependence in observed scaling relations:} We find no clear secondary dependence of $M_{\mathrm{mol}}$, $M_{\mathrm{mol}}/M_{\star}$, or $t_{\mathrm{dep}}$ on metallicity. This suggests that, once molecular gas is formed, its efficiency in forming stars is largely insensitive to metallicity.
    
    \item \textbf{Low-mass galaxies at cosmic noon follow established evolutionary trends:} At fixed stellar mass, galaxies at $z \sim 2$ have molecular-to-stellar mass ratios nearly an order of magnitude higher than local galaxies, while depletion times evolve only weakly (by a factor of $\sim 2$--$3$). These trends persist down to the lower-mass regime uniquely probed by ACE.
    
    \item \textbf{Refined fitting function for gas fractions:} We derive an empirical prescription for predicting of $M_{\mathrm{mol}}$ as a function of SFR and sSFR, with an intrinsic scatter of $\sigma \approx 0.2$ dex.
\end{itemize}

Overall, our results reinforce a unified picture in which star formation in galaxies is primarily governed by the availability of molecular gas, while variations in efficiency play a secondary role. The key scaling relations linking gas, stars, and star formation remain remarkably robust across nearly two orders of magnitude in stellar mass and over a wide range in metallicity, suggesting a common physical framework underlying galaxy evolution.

The ACE survey opens a critical new window onto the molecular gas properties of typical progenitors of present-day galaxies. Looking ahead, expanding such studies to more complete and less biased samples, particularly toward lower main-sequence offsets and even lower metallicities, will be essential for fully characterizing the gas cycle across cosmic time. Future observations targeting dense gas tracers, spatially resolved gas distributions, and molecular gas dynamics will further illuminate the physical mechanisms regulating star formation efficiency. A significant step forward, especially at the low-mass faint-CO regime, will require a significant increase in the line-sensitivity of ALMA, offered for example by the ALMA2040 concept \citep{Facchini2025}. Ultimately, these efforts will be key to developing a complete picture of galaxy evolution that consistently links gas accretion, star formation, and feedback across all cosmic epochs.

\begin{acknowledgements}
This paper makes use of the following ALMA data: ADS/JAO.ALMA\#2018.1.01128.S, 2024.1.00534.L. ALMA is a partnership of ESO (representing its member states), NSF (USA) and NINS (Japan), together with NRC (Canada), MOST and ASIAA (Taiwan), and KASI (Republic of Korea), in cooperation with the Republic of Chile. The Joint ALMA Observatory is operated by ESO, AUI/NRAO and NAOJ. We acknowledge assistance and computational support provided by Allegro, the European ALMA Regional Center node in the Netherlands. Some of the analysis presented herein builds on data products retrieved from the Dawn JWST Archive (DJA). DJA is an initiative of the Cosmic Dawn Center (DAWN), which is funded by the Danish National Research Foundation under grant DNRF140. This work has been funded by the European Research Council (ERC) under the European Union's Horizon 2020 research and innovation programme (DistantDust, Grant agreement No. 101117541). IL and IS acknowledge the grant Atracción de Talento Grant No. 2022-T1/TIC-20472, funded by the Comunidad de Madrid, Spain. MK acknowledges support from the Australian Research Council via the Discovery Early Career Researcher Award DE250100709. LAB. acknowledges support from the Dutch Research Council (NWO) under grant VI.Veni.242.055 (\url{https://doi.org/10.61686/LAJVP77714}). MP is funded by NASA grant ATP-23-0002. DN is grateful for support from NASA via grants ATP-21-0013 and ATP-23-0002. DL acknowledges the support from the Strategic Priority Research Program of the Chinese Academy of Sciences, grant No. XDB0800401.
\end{acknowledgements}

\bibliographystyle{aa}
\bibliography{mybib}

\clearpage
\onecolumn
\begin{appendix}
\section{Tabulated ACE galaxy properties}

In Table~\ref{tab:ace_properties} we list the properties of the ACE sample of galaxies used in this work.

\begin{table}[!htbp]
\centering
\caption{Overview of the properties of the ACE galaxies presented in this work. The listed properties are taken from \citet[for the CO flux density and molecular gas masses]{Langan2026}, and \citet{Shivaei2026} for the other properties, where we list the SFRs obtained from SED fitting. More details are given in Section~\ref{sec:data}.}
\label{tab:ace_properties}

\begin{tabular}{lcccccccc}
\hline\hline
ID & $z$ & $M_\ast$ & $\mathrm{SFR}$ & $12+\log(\mathrm{O/H})$ & $S_{\mathrm{CO(3{-}2)}}\Delta v$ & $M_\mathrm{mol}$ & $M_{\mathrm{mol}}/M_{\star}$ & $t_\mathrm{dep}$  \\
 & & $10^{10}\,\mathrm{M}_\odot$  & $\mathrm{M}_\odot\,\mathrm{yr}^{-1}$ &  & Jy km s$^{-1}$ & $10^{10}\,\mathrm{M}_\odot$ & &Gyr\\
\hline
3666 & $2.086$ & $0.16^{+0.04}_{-0.02}$ & $45.55 \pm 3.56$ & $8.29^{+0.02}_{-0.02}$ & $44.44 \pm 9.91$ & $3.31^{+0.88}_{-0.70}$ & $20.77^{+6.59}_{-5.80}$ & $0.73^{+0.21}_{-0.16}$ \\
24020 & $2.092$ & $0.89^{+0.15}_{-0.17}$ & $15.03 \pm 2.24$ & $8.32^{+0.03}_{-0.03}$ & $<32.90$ & $<2.20$ & $<2.79$ & $<1.59$ \\
5814 & $2.127$ & $2.19^{+0.27}_{-0.23}$ & $57.01 \pm 7.60$ & $8.45^{+0.02}_{-0.02}$ & $118.99 \pm 19.68$ & $4.68^{+0.93}_{-0.78}$ & $2.14^{+0.50}_{-0.41}$ & $0.82^{+0.21}_{-0.16}$ \\
6750 & $2.127$ & $0.47^{+0.06}_{-0.06}$ & $28.33 \pm 3.54$ & $8.35^{+0.02}_{-0.02}$ & $<34.86$ & $<2.17$ & $<5.01$ & $<0.83$ \\
13701 & $2.166$ & $1.32^{+0.18}_{-0.07}$ & $67.55 \pm 4.69$ & $8.55^{+0.02}_{-0.03}$ & $116.13 \pm 22.40$ & $3.35^{+0.80}_{-0.66}$ & $2.53^{+0.62}_{-0.57}$ & $0.50^{+0.13}_{-0.10}$ \\
13296 & $2.167$ & $1.81^{+0.22}_{-0.22}$ & $31.08 \pm 8.24$ & $8.61^{+0.05}_{-0.05}$ & $88.00 \pm 18.08$ & $1.83^{+0.57}_{-0.46}$ & $1.01^{+0.36}_{-0.27}$ & $0.59^{+0.30}_{-0.18}$ \\
5094 & $2.171$ & $2.13^{+0.77}_{-0.99}$ & $23.19 \pm 4.52$ & $8.47^{+0.05}_{-0.06}$ & $191.15 \pm 41.29$ & $6.97^{+2.23}_{-1.82}$ & $3.27^{+3.22}_{-1.15}$ & $3.01^{+1.27}_{-0.89}$ \\
25229 & $2.181$ & $1.59^{+0.28}_{-0.27}$ & $22.44 \pm 3.89$ & $8.29^{+0.03}_{-0.03}$ & $40.88 \pm 12.27$ & $2.82^{+1.06}_{-0.77}$ & $1.78^{+0.80}_{-0.54}$ & $1.26^{+0.56}_{-0.38}$ \\
19985 & $2.188$ & $2.83^{+0.35}_{-0.35}$ & $75.08 \pm 14.45$ & $8.31^{+0.01}_{-0.01}$ & $59.90 \pm 12.89$ & $4.23^{+1.04}_{-0.83}$ & $1.49^{+0.44}_{-0.33}$ & $0.56^{+0.20}_{-0.14}$ \\
6283 & $2.224$ & $0.71^{+0.16}_{-0.16}$ & $20.54 \pm 3.10$ & $8.22^{+0.02}_{-0.03}$ & $<36.92$ & $<4.08$ & $<6.83$ & $<2.18$ \\
16594 & $2.286$ & $0.92^{+0.19}_{-0.19}$ & $36.14 \pm 8.50$ & $8.45^{+0.03}_{-0.03}$ & $62.91 \pm 13.12$ & $2.67^{+0.72}_{-0.57}$ & $2.90^{+1.13}_{-0.76}$ & $0.74^{+0.32}_{-0.20}$ \\
2672 & $2.307$ & $4.08^{+1.18}_{-1.45}$ & $70.35 \pm 13.10$ & $8.38^{+0.05}_{-0.05}$ & $143.48 \pm 34.29$ & $8.07^{+2.99}_{-2.15}$ & $1.98^{+1.42}_{-0.65}$ & $1.15^{+0.52}_{-0.34}$ \\
3324 & $2.307$ & $1.99^{+0.41}_{-0.32}$ & $45.98 \pm 6.23$ & $8.57^{+0.05}_{-0.05}$ & $86.31 \pm 17.17$ & $2.49^{+0.75}_{-0.60}$ & $1.25^{+0.46}_{-0.35}$ & $0.54^{+0.19}_{-0.14}$ \\
3626 & $2.325$ & $1.47^{+0.15}_{-0.16}$ & $34.29 \pm 4.07$ & $8.19^{+0.03}_{-0.03}$ & $<33.17$ & $<4.06$ & $<2.91$ & $<1.26$ \\
5901 & $2.396$ & $0.53^{+0.16}_{-0.07}$ & $25.49 \pm 3.67$ & $8.45^{+0.02}_{-0.03}$ & $<36.76$ & $<1.95$ & $<3.97$ & $<0.84$ \\
9971 & $2.411$ & $2.56^{+0.53}_{-0.45}$ & $108.17 \pm 10.43$ & $8.36^{+0.02}_{-0.02}$ & $76.05 \pm 19.23$ & $4.98^{+1.50}_{-1.15}$ & $1.95^{+0.75}_{-0.54}$ & $0.46^{+0.15}_{-0.11}$ \\
9393 & $2.413$ & $0.78^{+0.20}_{-0.14}$ & $49.93 \pm 5.84$ & $8.37^{+0.02}_{-0.02}$ & $38.54 \pm 9.64$ & $2.58^{+0.77}_{-0.60}$ & $3.31^{+1.30}_{-0.98}$ & $0.52^{+0.17}_{-0.13}$ \\
3773 & $2.425$ & $0.44^{+0.08}_{-0.07}$ & $25.61 \pm 3.26$ & $8.27^{+0.03}_{-0.03}$ & $<37.88$ & $<4.22$ & $<10.59$ & $<1.75$ \\
4497 & $2.441$ & $1.79^{+0.44}_{-0.41}$ & $39.44 \pm 7.44$ & $8.51^{+0.04}_{-0.04}$ & $68.41 \pm 14.61$ & $2.61^{+0.76}_{-0.60}$ & $1.46^{+0.64}_{-0.42}$ & $0.66^{+0.26}_{-0.18}$ \\
8515 & $2.454$ & $0.13^{+0.06}_{-0.02}$ & $20.88 \pm 2.11$ & $8.27^{+0.02}_{-0.03}$ & $<41.06$ & $<4.83$ & $<40.90$ & $<2.43$ \\
19013 & $2.457$ & $0.75^{+0.12}_{-0.12}$ & $53.42 \pm 6.51$ & $8.52^{+0.03}_{-0.03}$ & $58.46 \pm 14.39$ & $2.26^{+0.70}_{-0.54}$ & $3.03^{+1.13}_{-0.82}$ & $0.42^{+0.15}_{-0.11}$ \\
24763 & $2.464$ & $0.90^{+0.14}_{-0.13}$ & $41.73 \pm 5.88$ & $8.45^{+0.05}_{-0.05}$ & $66.70 \pm 12.99$ & $3.59^{+1.14}_{-0.88}$ & $3.97^{+1.46}_{-1.08}$ & $0.86^{+0.32}_{-0.23}$ \\
22193 & $2.465$ & $0.78^{+0.33}_{-0.13}$ & $22.61 \pm 5.62$ & $8.44^{+0.04}_{-0.04}$ & $<33.82$ & $<2.08$ & $<2.90$ & $<1.09$ \\
19439 & $2.466$ & $0.56^{+0.11}_{-0.10}$ & $45.79 \pm 6.38$ & $8.22^{+0.02}_{-0.02}$ & $<31.74$ & $<4.23$ & $<8.67$ & $<1.01$ \\
21955 & $2.468$ & $0.42^{+0.10}_{-0.09}$ & $36.40 \pm 5.19$ & $8.50^{+0.03}_{-0.03}$ & $67.18 \pm 16.05$ & $2.88^{+0.87}_{-0.68}$ & $6.93^{+2.94}_{-2.02}$ & $0.79^{+0.28}_{-0.20}$ \\
8280 & $2.494$ & $3.90^{+0.20}_{-0.20}$ & $193.34 \pm 10.77$ & $8.51^{+0.05}_{-0.05}$ & $116.47 \pm 13.17$ & $4.46^{+1.03}_{-0.88}$ & $1.15^{+0.27}_{-0.23}$ & $0.23^{+0.06}_{-0.05}$ \\
\hline
\end{tabular}
\end{table}

\clearpage
\section{Literature compilation with metallicity information}
In Table~\ref{tab:literature_compilation_properties} we list the properties of the galaxies that are part of the literature compilation with metallicity and CO information.
\label{app:stacking-results}
\begin{table}[!htbp]
\centering
\caption{Properties of the galaxies included in the literature compilation with available gas-phase metallicity measurements (comprising the PHIBSS and ASPECS surveys). Metallicities for the ASPECS galaxies were calculated using the emission line fluxes tabulated in \citet{Kiyota2026}, whereas those for the PHIBSS targets were derived from strong-ionization line fluxes obtained by cross-matching the PHIBSS sample with JWST/NIRSpec observations within the DJA. Molecular hydrogen ($\mathrm{H}_2$) masses were computed by adopting the $\mathrm{CO}$ fluxes from \citet{Boogaard2019} for the ASPECS sources, and were drawn from the comprehensive literature compilation of \citet{Tacconi2020} for the PHIBSS sources. All listed molecular hydrogen masses are based on the \citet{Accurso2017} CO--to--\h2 conversion factor.  Stellar masses ($M_{\star}$) and star formation rates ($\mathrm{SFR}$) for both samples are based on SED fitting performed with the \textsc{Prospector} code. See Section~\ref{sec:data} and also \citet{Langan2026} for a description.}
\label{tab:literature_compilation_properties}
\small
\begin{tabular}{lcccccc}
\hline\hline
ID & $z$ & $M_{\star} / 10^{10}\,\mathrm{M}_\odot$ & $\mathrm{SFR}\,/\,\mathrm{M}_\odot\,\mathrm{yr}^{-1}$ & $12+\log(\mathrm{O/H})$ & CO line & $M_{\mathrm{mol}} / 10^{10}\,\mathrm{M}_\odot$ \\
\hline
GN4-7054 & $1.013$ & $23.51^{+1.27}_{-1.79}$ & $11.29 \pm 2.73$ & $8.82^{+0.02}_{-0.01}$ & CO(2--1) & $<1.68$ \\
EGS13017707 & $1.037$ & $5.55^{+2.88}_{-1.23}$ & $10.06 \pm 12.47$ & $8.57^{+0.03}_{-0.03}$ & CO(2--1) & $8.99^{+1.20}_{-1.12}$ \\
ASPECS/1mm.13 & $1.038$ & $4.32^{+0.35}_{-0.30}$ & $22.71 \pm 3.94$ & $8.65^{+0.01}_{-0.01}$ & CO(2--1) & $2.16^{+0.32}_{-0.28}$ \\
ASPECS/1mm.21 & $1.093$ & $6.65^{+0.58}_{-0.54}$ & $7.31 \pm 1.33$ & $8.64^{+0.04}_{-0.04}$ & CO(2--1) & $<0.66$ \\
ASPECS/1mm.16 & $1.317$ & $13.48^{+2.62}_{-2.44}$ & $0.48 \pm 1.52$ & $8.84^{+0.02}_{-0.03}$ & CO(2--1) & $0.98^{+0.94}_{-0.48}$ \\
ASPECS/1mm.14 & $1.997$ & $4.68^{+1.69}_{-2.88}$ & $2.70 \pm 9.23$ & $8.71^{+0.04}_{-0.04}$ & CO(1--0) & $<0.63$ \\
GN4-19913 & $2.013$ & $12.86^{+112.85}_{-1.66}$ & $815.63 \pm 939.69$ & $8.69^{+0.03}_{-0.03}$ & CO(3--2) & $<42.97$ \\
GN4-29743 & $2.187$ & $2.41^{+2.55}_{-0.66}$ & $61.00 \pm 38.64$ & $8.67^{+0.03}_{-0.03}$ & CO(3--2) & $2.60^{+1.10}_{-0.77}$ \\
zC406690 & $2.196$ & $6.11^{+11.73}_{-2.15}$ & $111.34 \pm 188.18$ & $8.69^{+0.08}_{-0.08}$ & CO(3--2) & $5.74^{+2.62}_{-1.85}$ \\
Q2343-BX610 & $2.211$ & $3.07^{+4.80}_{-1.11}$ & $423.91 \pm 242.84$ & $8.59^{+0.03}_{-0.03}$ & CO(3--2) & $25.32^{+3.54}_{-3.25}$ \\
ASPECS/1mm.04 & $2.454$ & $2.11^{+0.57}_{-0.42}$ & $57.49 \pm 7.14$ & $8.65^{+0.03}_{-0.03}$ & CO(3--2) & $6.55^{+0.86}_{-0.80}$ \\
ASPECS/1mm.01 & $2.543$ & $0.46^{+0.06}_{-0.05}$ & $226.62 \pm 11.56$ & $8.53^{+0.02}_{-0.02}$ & CO(3--2) & $30.21^{+1.98}_{-1.88}$ \\
ASPECS/1mm.07 & $2.581$ & $8.22^{+0.63}_{-0.55}$ & $26.67 \pm 5.41$ & $8.82^{+0.03}_{-0.03}$ & CO(3--2) & $1.87^{+0.50}_{-0.39}$ \\
ASPECS/1mm.06 & $2.696$ & $9.83^{+1.13}_{-1.31}$ & $133.16 \pm 16.25$ & $8.54^{+0.04}_{-0.04}$ & CO(3--2) & $23.75^{+4.50}_{-3.87}$ \\
\hline
\end{tabular}
\end{table}

\clearpage
\section{Stacking results}
In Table~\ref{tab:stacks} we list the galaxy properties, including CO(3--2) flux density and molecular gas mass, obtained for the respective stacked bins used in this work.
\begin{landscape}
\thispagestyle{empty}
\begin{table}[H]
    \centering
    \caption{Physical properties of the stacks used in this paper. A more detailed description of the stacking methodology, including error analysis, is presented in \citet{Geesink2026}.}
    \label{tab:stacks}
    \small

\begin{tabular}{@{}llcccccccc@{}}
\hline
Quantity & Bin & $z_{\mathrm{med}}$ & $S_{\mathrm{CO},\,3\text{-}2}\,\Delta v$ & $M_{\mathrm{H}_{2}}$ & $\log_{10}(M_{\star}/M_{\odot})$ & $\log_{10}(\mathrm{SFR}/M_{\odot}\,\mathrm{yr}^{-1})$ & $\log_{10}(\mathrm{sSFR}/\mathrm{yr}^{-1})$ & $12+\log_{10}(\mathrm{O/H})$ & $\log_{10}(\mathrm{SFR}/\mathrm{SFR}_{\mathrm{MS}})$ \\
 &  & & {\footnotesize (mJy\,km\,s$^{-1}$)} & {\footnotesize ($10^{10}\,M_{\odot}$)} &  \\
\hline
$\log_{10}(M_{\star}/M_{\odot})$ & < 9.80 & $2.43^{+0.01}_{-0.03}$ & $15.6^{+6.0}_{-5.4}$ & $1.37^{+0.53}_{-0.47}$ & $9.65^{+0.03}_{-0.02}$ & $1.45^{+0.05}_{-0.04}$ & $-8.14^{+0.07}_{-0.08}$ & $8.29^{+0.06}_{-0.01}$ & $0.40^{+0.05}_{-0.05}$ \\
 & 9.80 to 10.20 & $2.32^{+0.07}_{-0.04}$ & $28.4^{+8.5}_{-7.1}$ & $1.26^{+0.38}_{-0.32}$ & $9.95^{+0.01}_{-0.03}$ & $1.56^{+0.04}_{-0.03}$ & $-8.44^{+0.07}_{-0.10}$ & $8.44^{+0.01}_{-0.04}$ & $0.12^{+0.07}_{-0.09}$ \\
 & > 10.20 & $2.31^{+0.01}_{-0.12}$ & $85.7^{+9.1}_{-10.0}$ & $3.61^{+0.38}_{-0.53}$ & $10.34^{+0.07}_{-0.01}$ & $1.76^{+0.07}_{-0.05}$ & $-8.63^{+0.05}_{-0.01}$ & $8.45^{+0.03}_{-0.03}$ & $0.03^{+0.10}_{-0.01}$ \\
\addlinespace
$12+\log_{10}(\mathrm{O/H})$ & < 8.38 & $2.27^{+0.05}_{-0.05}$ & $17.1^{+6.8}_{-4.3}$ & $1.33^{+0.53}_{-0.34}$ & $9.89^{+0.03}_{-0.04}$ & $1.49^{+0.06}_{-0.04}$ & $-8.37^{+0.11}_{-0.08}$ & $8.29^{+0.01}_{-0.01}$ & $0.29^{+0.01}_{-0.10}$ \\
 & 8.38 to 8.48 & $2.38^{+0.04}_{-0.08}$ & $55.9^{+14.3}_{-14.1}$ & $2.55^{+0.66}_{-0.66}$ & $9.96^{+0.28}_{-0.04}$ & $1.58^{+0.08}_{-0.10}$ & $-8.44^{+0.08}_{-0.10}$ & $8.45^{+0.01}_{-0.01}$ & $0.17^{+0.04}_{-0.05}$ \\
 & > 8.48 & $2.37^{+0.07}_{-0.07}$ & $76.9^{+10.6}_{-11.9}$ & $2.63^{+0.39}_{-0.43}$ & $10.25^{+0.01}_{-0.13}$ & $1.63^{+0.07}_{-0.03}$ & $-8.41^{+0.10}_{-0.23}$ & $8.52^{+0.01}_{-0.01}$ & $0.19^{+0.16}_{-0.16}$ \\
\addlinespace
$\log_{10}(\mathrm{SFR}/M_{\odot}\,\mathrm{yr}^{-1})$ & <1.48 & $2.29^{+0.02}_{-0.09}$ & $13.7^{+7.4}_{-10.1}$ & $0.88^{+0.48}_{-0.65}$ & $9.87^{+0.05}_{-0.08}$ & $1.36^{+0.00}_{-0.01}$ & $-8.54^{+0.11}_{-0.11}$ & $8.33^{+0.04}_{-0.03}$ & $0.02^{+0.10}_{-0.05}$ \\
 & 1.48 to 1.69 & $2.37^{+0.06}_{-0.05}$ & $38.1^{+10.8}_{-9.4}$ & $1.74^{+0.50}_{-0.43}$ & $9.96^{+0.12}_{-0.10}$ & $1.61^{+0.01}_{-0.03}$ & $-8.37^{+0.11}_{-0.12}$ & $8.45^{+0.03}_{-0.04}$ & $0.20^{+0.09}_{-0.10}$ \\
 & >1.69 & $2.31^{+0.10}_{-0.00}$ & $76.7^{+12.0}_{-8.9}$ & $3.39^{+0.59}_{-0.39}$ & $10.34^{+0.07}_{-0.00}$ & $1.83^{+0.02}_{-0.00}$ & $-8.37^{+0.07}_{-0.03}$ & $8.45^{+0.01}_{-0.07}$ & $0.31^{+0.04}_{-0.13}$ \\
\addlinespace
$\log_{10}(\mathrm{sSFR}/\mathrm{yr}^{-1})$ & <-8.58 & $2.20^{+0.05}_{-0.03}$ & $49.2^{+9.1}_{-10.6}$ & $2.21^{+0.42}_{-0.48}$ & $10.26^{+0.02}_{-0.03}$ & $1.51^{+0.05}_{-0.08}$ & $-8.71^{+0.01}_{-0.05}$ & $8.41^{+0.04}_{-0.06}$ & $-0.02^{+0.02}_{-0.04}$ \\
 & -8.58 to 8.26 & $2.36^{+0.05}_{-0.05}$ & $52.7^{+11.9}_{-14.3}$ & $2.40^{+0.55}_{-0.66}$ & $10.05^{+0.17}_{-0.10}$ & $1.64^{+0.08}_{-0.07}$ & $-8.35^{+0.01}_{-0.04}$ & $8.45^{+0.01}_{-0.04}$ & $0.22^{+0.04}_{-0.02}$ \\
 & >-8.26 & $2.45^{+0.00}_{-0.04}$ & $24.7^{+10.7}_{-6.6}$ & $1.79^{+0.77}_{-0.48}$ & $9.67^{+0.07}_{-0.03}$ & $1.66^{+0.00}_{-0.10}$ & $-8.14^{+0.06}_{-0.00}$ & $8.35^{+0.02}_{-0.06}$ & $0.44^{+0.00}_{-0.04}$ \\
\addlinespace
$\log_{10}(\mathrm{SFR}/\mathrm{SFR}_{\mathrm{MS}})$ & <0.07 & $2.24^{+0.02}_{-0.04}$ & $42.4^{+10.4}_{-10.0}$ & $1.96^{+0.48}_{-0.47}$ & $10.23^{+0.03}_{-0.04}$ & $1.43^{+0.08}_{-0.07}$ & $-8.71^{+0.06}_{-0.05}$ & $8.41^{+0.04}_{-0.06}$ & $-0.02^{+0.01}_{-0.04}$ \\
 & 0.07 to 0.34 & $2.35^{+0.05}_{-0.05}$ & $46.4^{+11.5}_{-12.3}$ & $2.36^{+0.59}_{-0.64}$ & $9.96^{+0.20}_{-0.04}$ & $1.64^{+0.07}_{-0.06}$ & $-8.36^{+0.04}_{-0.03}$ & $8.41^{+0.03}_{-0.04}$ & $0.20^{+0.02}_{-0.01}$ \\
 & >0.34 & $2.45^{+0.01}_{-0.04}$ & $39.8^{+9.7}_{-8.8}$ & $2.74^{+0.67}_{-0.61}$ & $9.75^{+0.13}_{-0.07}$ & $1.66^{+0.01}_{-0.00}$ & $-8.14^{+0.06}_{-0.05}$ & $8.36^{+0.09}_{-0.01}$ & $0.44^{+0.01}_{-0.04}$ \\

\hline
\end{tabular}%

\end{table}

\end{landscape}
\clearpage
\end{appendix}

\end{document}